\documentclass[aps,showpacs]{revtex4}
\input{epsf}

\begin{document}

\title{\bf  Conditional Quantum Dynamics with Several Observers}
\author{ Jacek Dziarmaga$^{1,2}$, Diego A. R. Dalvit$^1$ 
         and 
         Wojciech H. Zurek$^1$ }
\affiliation{ 1) Los Alamos National Laboratory, Theoretical Division, 
             MS-B213, Los Alamos, New Mexico 87545 \\
          2) Instytut Fizyki Uniwersytetu Jagiello\'nskiego, 
             ul. Reymonta 4, 30-059 Krak\'ow, Poland
}

\begin{abstract} 

We consider several observers who monitor different parts of the
environment of a single quantum system and use their data to deduce its
state. We derive a set of conditional stochastic master equations that
describe the evolution of the density matrices each observer ascribes to
the system under the Markov approximation, and show that this problem can
be reduced to the case of a single ``super-observer", who has access to
all the acquired data. The key problem - consistency of the sets of data
acquired by different observers - is then reduced to the probability that
a given combination of data sets will be ever detected by the
``super-observer".  The resulting conditional master equations are applied
to several physical examples: homodyne detection of phonons in quantum
Brownian motion, photo-detection and homodyne detection of resonance
fluorescence from a two-level atom.  We introduce {\it relative purity} to
quantify the correlations between the information about the system
gathered by different observers from their measurements of the
environment. We find that observers gain the most information about the
state of the system and they agree the most about it when they measure the
environment observables with eigenstates most closely correlated with the
optimally predictable {\it pointer basis} of the system.

\end{abstract}


\maketitle


\section{ Introduction }

Information about the state of a quantum system is usually obtained not
from direct measurements on the system, but rather by monitoring its
environment \cite{whz93,whz98,whzadp}. Therefore, the environment is not
only a reservoir that selectively destroys quantum coherence, but also a
``communication channel" through which observers find out about the system
\cite{whz01}. The formalism that ascribes to the system a time-dependent
state deduced from a complete measurement of the environment - a quantum
trajectory - has been introduced some time ago
\cite{carmichael,wiseman,wisemanPhD,barchielli}.  Persistent monitoring of
a quantum system by the environment can single out a preferred set of
states, known as pointer states, which are the most robust in spite of the
interaction with the environment, that is, least perturbed by it
\cite{Zurek81,pointer}.  In \cite{ourPRL} we showed that under reasonable
assumptions pointer states remain the most robust even when a single
observer is performing continuous quantum measurement on a part of the
environment to extract information about the system. Here we use the
formalism of quantum trajectories
\cite{carmichael,wiseman,wisemanPhD,barchielli} to derive a set of
conditional master equations which describe density matrices inferred by
several observers simultaneously performing measurements on different
parts of the environment. The knowledge a given observer has about the
system, inferred from his measurement records, leads to his single
observer density matrix. One can also consider a super-observer with
access to all the records of all observers. His knowledge about the system
is encapsulated in a super-observer density matrix. Both density matrices
evolve according to conditional stochastic master equations that will be
derived in the next Section.

The paper is organized as follows. In Section II we derive conditional 
stochastic master equations for the situation where multiple 
observers monitoring the environment of a quantum system try to infer its 
state from their measurement data. Section III discusses correlations between
different measurement channels. In Section IV we apply the formalism 
to the quantum Brownian motion at zero temperature, a model  for which 
coherent states are perfect pointer states. Observers trying to infer
the state of the oscillator, initially prepared in a Schr\"dinger cat state
$(|z\rangle + |-z\rangle)/\sqrt{2}$ of large amplitude coherent states,
eventually fully agree when they measure in a basis correlated to the pointer
states. In Sections V and VI we consider the model 
of resonance fluorescence of a two-level atom, for which pointer states 
do not really exist, and even the most predictable states are quite unpredictable.
We discuss correlations between single observer state 
assignments for different measurement schemes, such as photo-detection and 
homodyne detection. Finally, Section VII contains our conclusions.


\section{Conditional Master Equations for Several Observers}

  Imagine a system $\cal S$ coupled to an environment $\cal E$. The state
of the environment is monitored by a set of detectors
$\{ {\cal D}_i \} \;  (i=1, \ldots, C)$ in measurement basis that can be 
different 
for different detectors. When the results of these measurements are 
ignored (which in technical terms corresponds to tracing over $\cal E$ as 
well as over all the records of $\{ {\cal D}_i \}$), the
reduced density matrix of the system $\varrho(t)$ evolves according to an
``unconditional'' master equation (UME)

\begin{equation}
d\varrho(t)=dt{\cal L}\odot\varrho(t)\;\;,
\label{umeuno}
\end{equation}
within an infinitesimal time step $dt$. Here ${\cal L}$ is a linear 
super-operator acting on $\varrho$. This master equation is ``unconditional'' 
in the sense that all the information about the records of 
$\{ {\cal D}_i \}$ has been ignored. Note that we are assuming that 
this evolution is Markovian, so that the state of the system at time 
$t+dt$ only 
depends on its state at time $t$. In what follows we shall make a still stronger 
assumption that the state of the fragments of the environment prior to the 
interaction with the system is always the same. 
In particular, the fragments of the environment that 
have  interacted with ${\cal S}$ do not interact with it again. This quantum 
Markov approximation is accurate in a typical scattering situation. Then there
is a natural distinction between input and output fields. The input field is 
assumed to be always described by the same density matrix $\mu$. The output 
field evolves in a way it must to account for its outward propagation. It is 
entangled with the system, and is eventually measured 
sufficiently far from the system so that the detection does not disturb 
the system. A spontaneous emission from, say, a two-level atom in free space 
is a particular example of the scattering situation where the input density 
matrix $\mu$ is a vacuum state.


\subsection{Multiple measurement channels}

  Suppose that the measurement records are not ignored but, instead, as is typically
the case, they
are used to extract information about $\cal S$. Let us further assume that
$\cal E$ includes parts $ \{ {\cal E}_i \}$ numbered by the index 
$i ~ (i=1, \ldots, C)$ 
and that such parts are coupled to the detectors $\{ {\cal D}_i \}$. 
These detectors correspond to different measurement channels. We model the 
measurements by detector ${\cal D}_i$ as a projection of the detector's 
state in a measurement basis with outcome $dN_i(t)$. In the example of
spontaneous emission the parts $\{ {\cal E}_i \}$ can be chosen as
different directions of photon emission monitored by different detectors
$\{ {\cal D}_i \}$, while $dN_i(t)$ may be the number of photons collected
by the detector $i$ during the infinitesimal time interval from $t$ to $t+dt$. 
No matter how much the measurement is delayed or how 
long it takes to decohere the state of the detector ${\cal D}_i$, the 
eventual outcome $dN_i(t)$ affects the knowledge about the state of the 
system at the time $t$ when the environment part ${\cal E}_i$ got 
entangled with $\cal S$. In general the measurement basis can be 
non-orthogonal and/or over-complete, but in the following, for the 
sake of simplicity, we restrict ourselves to the complete orthogonal case.

We assume that at $t=0$ the state of $\cal S$ was a density matrix 
$\varrho(0)$. As time goes on, at every time step $dt$ the system $\cal S$ 
is getting entangled with a new fragment of the environment $\cal E$. 
After interaction with $\cal S$ some of these fragments are not detected, 
while some other fragments, $\{ {\cal E}_i \}$, are measured by the 
detectors $\{ {\cal D}_i \}$. After time $t$ the initially 
uncorrelated state of $\cal S+E+D$ evolves into an entangled state 
$\rho_{\cal SED}(t)$ (by $\cal D$ we are denoting the set of detectors
$\{ {\cal D}_i \}$). Ignoring the state of $\cal ED$ we get the 
unconditional density matrix

\begin{equation}
\varrho(t)~=~{\rm Tr}_{\cal ED}~\rho_{\cal SED}(t) ,
\label{varrho}
\end{equation}
which evolves according to the UME, Eq. (\ref{umeuno}). Let us call 
${\cal N}_{\alpha}(t)=(N_1(t), \ldots, N_C(t))$ the set of strings of multiple
channels measurement records. That is, ${\cal N}_{\alpha}(t)$ 
is a particular history of measurement results $\{ dN_i(t') \}$ on all 
channels till time $t$, and the subscript $\alpha$ denotes a particular
set of records. Then, the unconditional
density matrix $\varrho(t)$ is a sum over all possible sets 
${\cal N}_{\alpha}(t)$,  namely

\begin{equation}
\varrho(t)=~
\sum_{ {\cal N}_{\alpha}(t) }~
{\rm Tr}_{\cal ED}~
{\cal P}_{ {\cal N}_{\alpha}(t) }
~\rho_{\cal SED}(t) .
\end{equation}
Here ${\cal P}_{ {\cal N}_{\alpha}(t) } = {\cal P}_{N_1(t)} \otimes \ldots 
\otimes {\cal P}_{N_C(t)}$ projects the state of all detectors
according to the measurement
records $N_i(t) (i=1, \ldots, C)$. Different branches of the wave-function of 
$\cal SED$ are labeled by different sets ${\cal N}_{\alpha}(t)$ of
possible measurement records. A branch 
${\cal N}_{\alpha}(t)$ has a probability
  
\begin{equation}
p[{\cal N}_{\alpha}(t) ]~=~
{\rm Tr}_{\cal SED}~{\cal P}_{{\cal N}_{\alpha}(t)}~\rho_{\cal SED}(t)~.
\label{pset}
\end{equation}
 
A hypothetical ``super-observer'', who knows all the measurement 
records of 
all detectors,
${\cal N}_{\alpha}(t)=(N_1(t), \ldots, N_C(t))$, 
ascribes to $\cal S$ a state given by a multiple observers (or super-observer)
conditional density matrix

\begin{equation}
\rho[ {\cal N}_{\alpha}(t)]~=~
\frac{ {\rm Tr}_{\cal ED}~{\cal P}_{{\cal N}_{\alpha}(t)}~\rho_{\cal SED}(t) }
     { {\rm Tr}_{\cal S}~
       {\rm Tr}_{\cal ED}~{\cal P}_{{\cal N}_{\alpha}(t)}~\rho_{\cal SED}(t) }
=
\frac{ {\rm Tr}_{\cal ED}~{\cal P}_{{\cal N}_{\alpha}(t)}~\rho_{\cal SED}(t) }
	 { p[{\cal N}_{\alpha}(t)]} ,
\label{ren}
\end{equation}
normalized so that ${\rm Tr}_{\cal S}\rho[{\cal N}_{\alpha}(t)]=1$. From the 
point of view of the super-observer the set of records ${\cal N}_{\alpha}(t)$ 
actually 
happens. His description is necessarily probabilistic: at $t=0$ he could only 
calculate probabilities of different outcomes (Eq. (\ref{pset})), but could 
not predict his actual set of outcomes ${\cal N}_{\alpha}(t)$.

We assign to each measurement channel $i$ an observer $i$ who knows 
only his own record $N_i(t)$. He ascribes to $\cal S$ a state given 
by a single observer conditional density matrix

\begin{equation}
\rho[N_i(t)]~=~
\frac{ {\rm Tr}_{\cal ED}~{\cal P}_{N_i(t)}~\rho_{\cal SED}(t) }
     { {\rm Tr}_{\cal S}~
       {\rm Tr}_{\cal ED}~{\cal P}_{N_i(t)}~\rho_{\cal SED}(t) } ,
\label{rensingle}
\end{equation}
conditioned only on his own record $N_i(t)$. It is easy to check that the
single observer density matrix is an average over the
super-observer density matrices with all records that are not
known to the observer $i$. Indeed, denoting
${\cal N}_{\beta}^{N_i(t)} = (N_1(t), \ldots, N_i(t), \ldots, N_C(t))$ 
a set of strings 
of multiple channel records that contain the particular record $N_i(t)$
in channel $i$, we have

\begin{equation}
\rho[N_i(t)]~=~
\sum_{ {\cal N}_{\beta}^{N_i(t)} } ~
\frac{ p[ {\cal N}_{\beta}^{N_i(t)} ]~\rho[ {\cal N}_{\beta}^{N_i(t)}] }
{p[N_i(t)]} .
\label{singlechannel}
\end{equation}
The probability distribution for the measurement record $N_i(t)$ is given by

\begin{equation}
p[N_i(t)] = 
{\rm Tr}_{\cal SED}~ {\cal P}_{N_i(t)}~ \rho_{\cal SED}(t) =
\sum_{ {\cal N}_{\beta}^{N_i(t)} } p[ {\cal N}_{\beta}^{N_i(t)} ] ,
\label{singlechannelprob}
\end{equation}
which is analogous to Eq. (\ref{pset}).

The issue of compatibility of density matrices ascribed to a system by 
different observers
was first considered by Peierls \cite{peierls}. He noted that the state assignments
of various observers cannot be arbitrarily different, and proposed that, in order to avoid contradiction between 
different state assignments,
the product of the corresponding density matrices should be non-zero. A 
second condition put forward
in \cite{peierls}, namely that the different density matrices should 
commute, was later shown
to be too restrictive \cite{fuchs}. Necessary and sufficient condition 
for compatibility of
several density matrices turns out to be simple \cite{mermin}: Their 
supports must share at least
one state (the support of a density matrix is the subspace spanned by all 
its eigenvectors with non
zero eigenvalues). 

In our setting  the issue of compatibility of various 
single observer density matrices
$\rho[N_i(t)]$ is settled very naturally. Indeed, the probability 
$p[ {\cal N}_{\alpha}(t) ]$ defined
in Eq. (\ref{pset}) provides a measure of compatibility of different sets of
outcomes: $p[ {\cal N}_{\alpha}(t)]=0$ when records from different channels 
in the set ${\cal N}_{\alpha}(t)$ are mutually contradictory. It is easy to see that 
this condition is equivalent to the pre-requisite compatibility, i.e. the overlap of
support \cite{peierls,fuchs,mermin}. Moreover, $p[ {\cal N}_{\alpha}(t)]$ quantifies 
this compatibility, at least in the multiple observer setting we are about to investigate
in more detail.


\subsection{Conditional master equations}

  The evolution of the unconditional density matrix 
$\varrho(t)$ is determined by
the unconditional master equation, Eq. (\ref{umeuno}). 
We now derive master equations that describe the evolution of the 
super-observer 
density matrix $\rho[{\cal N}_{\alpha}(t)]$,  and of the density matrix 
$\rho[N_i(t)]$ of observer $i$,  conditioned on their respective
measurement results. Within the Markovian approximation, 
the master equation for the single observer conditional density matrix has 
the form \cite{carmichael,wisemanPhD}

\begin{equation}
d\rho[{\cal N}_{\alpha}(t)]=
dt{\cal L}\odot\rho[{\cal N}_{\alpha}(t)]+
{\cal M}_{d{\cal N}_{\alpha}(t)} \odot\rho[{\cal N}_{\alpha}(t)]   \;\;.
\label{MOCMEuno}
\end{equation}
In our case of the super-observer that ``single observer'' has access to all
measurement records ${\cal N}_{\alpha}(t)$. 
The super-operator ${\cal M}_{d{\cal N}_{\alpha}(t)}$ 
conditions $\rho[{\cal N}_{\alpha}(t)]$ on 
present measurement results 
$d{\cal N}_{\alpha}(t)=(dN_1(t), \ldots, dN_C(t))$ 
that are obtained at time $t$. 
It can be written as a sum of super-operators that depend on the measurement 
results on individual channels,  
${\cal M}_{d{\cal N}_{\alpha}(t)}= {\cal M}_{dN_1(t)} + \ldots +
{\cal M}_{dN_C(t)}$. Each 
super-operator ${\cal M}_{dN_i(t)}$ takes density operators to density
operators and depends on the particular measurement strategy 
implemented in the measurement channel $i$.
Examples of measurement strategies are point processes, such
as photo-counting of optical fields, and diffusive processes, 
such as homodyne or heterodyne detection of optical fields.
Also, ${\cal M}_{d{\cal N}_{\alpha}(t)} \odot\rho[{\cal N}_{\alpha}(t)]$ 
is nonlinear in 
$\rho[{\cal N}_{\alpha}(t)]$ and linear in the set of measurement results 
$d{\cal N}_{\alpha}(t)$. 
The nonlinearity comes into play in Eq. (\ref{ren}) when we
normalize the density matrix. The action of this  super-operator on 
$\rho[{\cal N}_{\alpha}(t)]$ is a generalization of the apparent 
``collapse of the wave-function" experienced by the super-observer 
confined to the branch ${\cal N}_{\alpha}(t)$. 
For a derivation of the super-operators ${\cal M}_{d N_i(t)}$
in terms of the projectors ${\cal P}_{N_i(t)}$ see the formalism of operations
and effects described in \cite{barginsky,gardiner}.

When measurement results $d{\cal N}_{\alpha}(t)$ are ignored, $\varrho(t)$ 
should follow the UME (Eq. (\ref{umeuno})). In other words, the sum of the 
super-operator 
${\cal M}_{d{\cal N}_{\alpha}(t)}$ over all possible strings of measurement 
results 
$d{\cal N}_{\alpha}(t)$ should vanish, that is:  

\begin{equation}
\sum_{ d{\cal N}_{\alpha}(t) } 
{\cal M}_{d{\cal N}_{\alpha}(t)}  \odot\rho[{\cal N}_{\alpha}(t)]=~0~.
\end{equation}

We can also write down the master equation for the density matrix 
$\rho[N_i(t)]$ that observer $i$, who knows
only his own records $N_i(t)$, ascribes to the system ${\cal S}$. Using 
Eqs.(\ref{singlechannel},\ref{MOCMEuno}) we obtain

\begin{equation}
d \rho[N_i(t)] = dt {\cal L} \odot \rho[N_i(t)] +
{\cal M}_{dN_i(t)} \odot \rho[N_i(t)] .
\end{equation}
Indeed, this equation has the same form as that of the super-observer 
Eq. (\ref{MOCMEuno}). 
The super-operator ${\cal M}_{dN_i(t)}$ depends only on the measurement 
result $dN_i(t)$ 
of observer $i$, and it is defined as an average over all records of 
all other observers 
unknown to him,

\begin{equation}
{\cal M}_{dN_i(t)} \odot \rho[N_i(t)] = 
\sum_{{\cal N}_{\beta}^{N_i(t)}} 
\frac{p[{\cal N}_{\beta}^{N_i(t)}]}{p[N_i(t)]} ~
{\cal M}_{d{\cal N}_{\beta}^{dN_i(t)}} \odot
\rho[{\cal N}_{\beta}^{N_i(t)}]
 ,
\end{equation}
where $d{\cal N}_{\beta}^{dN_i(t)}$ is any string of multiple channel  
measurement
record that contains the particular measurement results $dN_i(t)$ on 
channel $i$.


\section{Correlations between different measurement channels}

We study correlations between measurement records on 
different measurement
channels, say channels $i$ and $j$. It is clear that in order to do so 
it is necessary 
to compare corresponding records $N_i(t)$ and $N_j(t)$. Therefore, we 
imagine there is someone
who has access to the records on both channels, or the two observers 
with access to channels
$i$ and $j$ communicate with each other and share their measurement 
records. Whatever the
case is, we can think that there is a (super)observer who has access 
to the two 
measurement channels and whose string of records is $(N_i(t), N_j(t))$. 
To follow 
the line of thought of previous sections, we will instead consider the 
super-observer
who has access to all measurement channels, and whose string of records 
contain the particular
record $N_i(t)$ on channel $i$, and the particular record $N_j(t)$ on 
channel $j$, i.e.,
his string of records is 
${\cal N}_{\gamma}^{N_i(t), N_j(t)}=(N_1(t), \ldots, N_i(t), 
\ldots, N_j(t), \ldots, N_C(t))$.
Here the subscript $\gamma$ denotes a particular set of multiple 
channel records that contains
records $N_i(t)$ and $N_j(t)$ in channels $i$ and $j$, respectively.

We define the average relative purity between the states ascribed 
to ${\cal S}$ by two such 
observers $i$ and $j$ as

\begin{eqnarray}
O_{i j}(t) &=& \sum_{N_i(t), N_j(t)} 
\sum_{{\cal N}_{\gamma}^{N_i(t), N_j(t)}}
p[ {\cal N}_{\gamma}^{N_i(t), N_j(t)}]
{\rm Tr}_{{\cal S}} \rho[N_i(t)] ~ \rho[N_j(t)] \nonumber \\
&=&
\sum_{N_i(t), N_j(t)} p[N_i(t), N_j(t)] {\rm Tr}_{\cal S} \rho[N_i(t)] ~ 
\rho[N_j(t)] ,
\label{Oij}
\end{eqnarray}
where $p[N_i(t), N_j(t)]$ is the joint probability distribution for records
$N_i(t)$ and $N_j(t)$, given by

\begin{equation}
p[N_i(t), N_j(t)] = \sum_{{\cal N}_{\gamma}^{N_i(t), N_j(t)}} 
p[ {\cal N}_{\gamma}^{N_i(t), N_j(t)} ] .
\end{equation}
We also introduce the average relative purity $O_i(t)$ 
between the states of ${\cal S}$
ascribed by the observer $i$ and the super-observer, whose respective
measurement records are $N_i(t)$ and $ {\cal N}_{\beta}^{N_i(t)}$,
i.e., the particular record $N_i(t)$ on channel $i$ is contained in the 
super-observer string of multiple channel records,
${\cal N}_{\beta}^{N_i(t)} = (N_1(t), \ldots, N_i(t), \ldots, N_C(t))$.
Here $\beta$ denotes a particular set of super-observer records that contains
the record $N_i(t)$ in channel $i$. We define

\begin{equation}
O_i(t) =
\sum_{N_i(t)} \sum_{{\cal N}_{\beta}^{N_i(t)}}
p[{\cal N}_{\beta}^{N_i(t)}]
{\rm Tr}_{{\cal S}} \rho[N_i(t)] ~ \rho[{\cal N}_{\beta}^{N_i(t)}] .
\label{Oisuper}
\end{equation}
Using Eqs.(\ref{singlechannel}, \ref{singlechannelprob}) 
it is easy to check that this average relative purity 
equals the average purity of the state ascribed to ${\cal S}$ by
observer $i$, and that it is also equal to the autocorrelation $O_{i i}(t)$ introduced
in Eq. (\ref{Oij}), namely

\begin{equation}
O_i(t) = O_{i i}(t) =
\sum_{N_i(t)} p[N_i(t)] {\rm Tr}_{{\cal S}} \rho^2[N_i(t)] .
\label{Oii}
\end{equation}
In other words, an observer $i$ has no more information about the 
super-observer's records than the information already contained in his own 
records. 

A better measure of correlations between density matrices is fidelity 
\cite{fidelity}, 
which is defined as 
$F(\rho_i, \rho_j) = \{ {\rm Tr}[ (\sqrt{\rho_i} \rho_j \sqrt{\rho_i})^{1/2} ]
\}^2$.
This can be easily calculated in two dimensions:
$F(\rho_i, \rho_j) = {\rm Tr}(\rho_i \rho_j) + 2 ({\rm det} \rho_i {\rm det} 
\rho_j)^{1/2}$.
Unfortunately it is more
difficult to compute for more general cases, and for this reason we will use 
in the following 
relative purity as a measure of correlations.

In order to make a quantative study of correlations between different 
measurement channels
it will be convenient to work in the framework of stochastic differential 
master equations.
We first simplify our notation: we will denote the density matrix ascribed 
to the system by 
the super-observer, who has measurement records ${\cal N}_{\beta}^{N_i(t)}$, as

\begin{equation}
\rho[{\cal N}_{\beta}^{N_i(t)}] \equiv \rho(t), 
\end{equation}
and the density matrix of observer $i$, whose measurement record is $N_i(t)$, 
as

\begin{equation}
\rho[N_i(t)] \equiv \rho_i(t) .
\end{equation}
Also, we will denote the measurement records ${\cal N}_{\beta}^{N_i(t)}$ 
of the super-observer, and his measurement results 
$d{\cal N}_{\beta}^{dN_i(t)}$, that 
respectively contain the measurement record $N_i(t)$ and the measurement 
result 
$dN_i(t)$ on channel $i$, as

\begin{eqnarray}
{\cal N}_{\beta}^{dN_i(t)} & \equiv & {\cal N}(t) , \nonumber \\
d{\cal N}_{\beta}^{dN_i(t)} & \equiv & d{\cal N}(t) .
\end{eqnarray}
We emphasize again that, in the study of correlations, we consider the 
situation
when the single channel measurement record $N_i(t)$ is contained
in the super-observer records ${\cal N}_{\beta}={\cal N}_{\beta}^{N_i(t)}$. 
Similarly,
when we study correlations between measurement records on two different 
channels $i$ and
$j$, we consider the situation when the two single channel measurement records
$N_i(t)$ and $N_j(t)$ are both contained in the super-observer records 
${\cal N}_{\beta} = {\cal N}_{\beta}^{N_i(t), N_j(t)}$.

  The conditional master equation for the super-observer, that conditions his 
$\rho(t)$ 
on current measurement results $d{\cal N}(t)$, can be upgraded to a 
{\it stochastic 
master equation} (SME). A SME is a conditional master equation
plus a probability distribution $P(d{\cal N}(t))$ for the measurement
results $d{\cal N}(t)$. This probability distribution can be obtained from 
Eq. (\ref{pset}). $P[d{\cal N}(t)]$ is a conditional probability to get current
measurement results $d{\cal N}(t)$ provided that the measurement records of 
the super-observer
until time $t$ are ${\cal N}(t)$, that is

\begin{equation}
P[d {\cal N}(t) ] \equiv  p[ d{\cal N}(t) | {\cal N}(t) ] =
\frac{  p[ d{\cal N}(t) , {\cal N}(t)]  }{ p[ {\cal N}(t)] } ,
\end{equation}
where we used Bayes rule. Here $p[d{\cal N}(t), {\cal N}(t)]$ is the joint 
probability
of having measurement records ${\cal N}(t)$ until time $t$, and of having 
measurement results
$d{\cal N}(t)$ at time $t$. In the Markovian approximation $P[d{\cal N}(t)]$ 
depends on the records ${\cal N}(t)$ through the conditional density matrix 
$\rho(t)$,

\begin{equation}
P[d{\cal N}(t)] = P[d{\cal N}(t) | \rho(t)] .
\label{Puno}
\end{equation}
The dependence of this probability distribution on the super-observer 
$\rho(t)$ leads to 
correlations between different measurement channels we are going to explore 
using a set
of SMEs describing the stochastic evolutions for $\rho(t)$ and $\rho_i(t)$.
This set of stochastic master equations is given by

\begin{eqnarray}
d\rho(t) &=& dt{\cal L}\odot\rho(t)+ {\cal M}_{d{\cal N}(t)}\odot\rho(t) , \\
d\rho_i(t) &=& dt{\cal L}\odot\rho_i(t)+ {\cal M}_{dN_i(t)}\odot\rho_i(t) , \\
P[dN_1(t), \ldots , dN_C(t) | \rho(t)] &=&
\frac{
{\rm Tr}_{\cal SED}~{\cal P}_{{\cal N}_{\alpha}(t), d{\cal N}_{\alpha}(t)}
~\rho_{\cal SED}(t+dt)
}
{ {\rm Tr}_{\cal SED}~{\cal P}_{{\cal N}_{\alpha}(t)}
~\rho_{\cal SED}(t)} .
\label{MCSME}
\end{eqnarray} 
We refer to this set of equations as a multiple channels stochastic
master equation (MCSME).

We will use the above formalism to address questions regarding correlations 
between 
measurements on different channels:

\begin{itemize}

\item What is the average correlation between the density matrix of a
single observer $\rho_i$ and the super-observer's density matrix $\rho$? 
We shall quantify this correlation by the average relative purity 
$O_i(t)$ defined in Eq. (\ref{Oisuper}). As a short-hand, we will write it as
$O_i(t) = \overline{{\rm Tr} \rho_i(t) \rho(t)}$, where the overline means the
weighted average defined in Eq. (\ref{Oisuper}). This relative purity is a 
measure of how different, on average, is the knowledge of the observer $i$ from the 
knowledge he would have 
had if he had access to the records of all the other observers. He cannot 
know more about 
the state $\rho(t)$ that the super-observer ascribes to the system that he 
can infer from his 
own measurement record only. The extracted information can be 
measured by the average purity of the single observer state 
$O_{i i}(t)= \overline{{\rm Tr} \rho_i^2(t)}$ (Eq. (\ref{Oii})). 
Even if the super-observer density matrix $\rho(t)$ had higher average
purity, the average relative purity $O_i(t)$ would be equal to the single 
observer
purity, $O_i(t)=O_{i i}(t)$, as derived in Eq. (\ref{Oii}). 
This equality will be illustrated with several examples in Sections V and VI. 
$O_{i i}(t)$ is maximal for measurements in a basis correlated with the 
pointer states
\cite{ourPRL}. 
%

\item What is the average correlation between different single observer
density matrices $\rho_i$ and $\rho_j$? 
We shall quantify this correlation by the average relative
purity $O_{i j}(t)$ defined in Eq. (\ref{Oij}), that as a short-hand we write as
$O_{i j}(t) =\overline{ {\rm Tr} \rho_i(t) \rho_j(t)}$.
In other words, how much do different observers agree about the state of the 
system? 
In Section IV we will show that for an initial Schr\"odinger cat state made of
large amplitude coherent states (coherent states are perfect pointer states
in the model of zero temperature quantum Brownian motion), and for measurements
in a basis of the environment correlated with them, observers will, 
after an initial transient, reach full agreement,
$O_{i j}(t\rightarrow\infty)=1$.  Typically, as
seen in the examples of Sections V and VI, the agreement is not perfect
but it gets better when the observers' measurement basis get closer to
those environmental states correlated to the pointer basis of the system. 
For resonance fluorescence from a
two-level atom subjected to direct photo-detection (see Section V) we find
an anti-correlation, $O_{i j} < 1/2$. Each observer learns
something about the state of the system but their estimates of the state
$\rho_i(t)$ are anti-correlated. The two-level atom is very far from
being classical and, what is more, photo-detection is very far from being a
measurement in a basis correlated with the most predictable states.

\end{itemize} 


\section{ Quantum Brownian Motion }

\subsection{Correlation of the outcomes for pointer state measurements}

  In this section we consider the well-known model of quantum Brownian
motion consisting of a harmonic oscillator (the system) interacting with a
reservoir of harmonic oscillators (phonons) with a position-position
coupling. We will restrict ourselves to the case of a zero temperature
environment. This model represents a damped harmonic oscillator. The
self-Hamiltonian for the system is $H=\omega a^{\dagger} a$, where
$\omega$ is the frequency of the oscillator and $a,a^{\dagger}$ are
bosonic annihilation/creation operators. Imagine that a set of observers
perform homodyne detection measurements on the environment of phonons so
that each of them gains information about the state of the system
oscillator. Given the set of records of all those observers, the MCSME of
the system oscillator is \footnote{In Appendix A we derive the MCSME for
the model of resonance fluorescence from a two-level atom. From that
starting point it is straightforward to get the corresponding master
equation for zero temperature quantum Brownian motion.}

\begin{eqnarray}\label{QBMCMEO}
&&
d\rho = 
dt \; \left( a \rho a^{\dagger} -
\frac{1}{2} a^{\dagger}a \rho - 
                     \frac{1}{2} \rho a^{\dagger}a  \right) +
\sum_i
     \left( dN_i-\overline{dN_i(\rho)} \right) \;
     \left( \frac{   (a+\gamma)\rho 
(a^{\dagger}+\gamma^{\star})}  
     { {\rm Tr}[ (a+\gamma)\rho (a^{\dagger}+\gamma^{\star}) ] }
                 -\rho  \right) ,
\label{QBM_MOCME}\\
&&
d\rho_i = 
dt \; \left( a \rho_i a^{\dagger} -
             \frac{1}{2} a^{\dagger}a \rho_i -
             \frac{1}{2} \rho_i a^{\dagger}a  \right) +
\left(dN_i -\overline{dN_i(\rho_i)}\right)
\left( \frac{   (a+\gamma)\rho_i (a^{\dagger}+\gamma^{\star}) }
            { {\rm Tr}[(a+\gamma)\rho_i(a^{\dagger}+\gamma^{\star})]}
                 -\rho_i
          \right) ,
\label{QBM_SOCME} \\
&&
\overline{dN_i(\rho)} =                       
\eta_i dt [ R^2 + R e^{-i\phi} {\rm Tr} \rho a +  
                   Re^{+i\phi} {\rm Tr} \rho a^{\dagger}+  
                   {\rm Tr}\rho a^{\dagger}a              ]\;. 
\label{QBM_dN}
\end{eqnarray}
These equations for the conditional evolution of the density matrices of
the system, written in the interaction picture representation, are valid
in the rotating wave approximation. Here we use It\^{o} version of
stochastic calculus. The first terms on the RHS of
Eqs. (\ref{QBM_MOCME},\ref{QBM_SOCME}) are of Lindblad form and describe
damping and decoherence due to spontaneous emission of phonons. We have
set the damping coefficient to one. The second (stochastic) terms feed
back into the master equation information about the state of the system
gained by observers.  The coefficient $\gamma=R\exp(i\phi)$ is the
amplitude of the local oscillator in the homodyne detector \cite{qoptics}.
For simplicity, we are asumming here that all observers perform the same
kind of homodyne detection, so that the amplitudes $R_i$ and phases
$\phi_i$ are all equal. We will lift this restriction in later examples.  
The number of phonons detected by observer $i$ in an infinitesimal
interval from $t$ to $t+dt$ is $dN_i(t)\in\{0,1\}$, with an average given
by Eq. (\ref{QBM_dN})  and $\overline{dN_i dN_j}= \delta_{i j} \;
\overline{dN_i}$.  The efficiencies $\eta_i$ of different detectors can be
defined as the fractions of phonons monitored by particular detectors. In
the phonodetection limit ($R=0$), the average detection rate Eq.
(\ref{QBM_dN}) is proportional to the average occupation number. Whenever
a phonon is detected ($dN_i=1$ for any $i$) the occupation numbers in
$\rho$ are reduced by one. In the homodyne limit ($R\gg 1$) the detection
rates measure the coherent amplitude ${\rm Tr}\rho
(e^{+i\phi}a+e^{-i\phi}a^{\dagger})$ of the state of the system.

  To illustrate how different observers are gaining information about the
system and how correlations between different measurement channels arise
in the process of continuous measurement we consider superpositions of
large amplitude coherent states. According to the exact solution \cite{exact}
coherent states $|z_0\rangle$, such that $a|z_0\rangle=z_0|z_0\rangle$,
decay to the ground state like $|z_0~e^{-t}\rangle=|z\rangle$ without
producing any entropy. At $T=0$ they are the perfect pointer states of the
quantum Brownian motion model \cite{zhp}. The decay to the ground state
takes place on a time scale of the order of the damping rate, which we
have set to 1. In a subspace spanned by $|+z\rangle$ and
$|-z\rangle$ a general density matrix is

\begin{equation}
\rho(t)=
\frac{1+A(t)}{2} |+z\rangle\langle +z| +
\frac{1-A(t)}{2} |-z\rangle\langle -z| +
C(t) |+z\rangle\langle -z| +
C^*(t) |-z\rangle\langle +z|~.
\label{AC}
\end{equation}
Substitution of this density matrix into Eq. (\ref{QBM_MOCME}), and
subsequent left and right projections on $|\pm z\rangle$ \footnote{In the
limit $r\gg 1$, the states $|+z\rangle$ and $|-z \rangle$ are
approximately orthogonal.} give stochastic differential equations for
$A(t)$ and $C(t)$. $C(t)$ decays to $0$ on a decoherence timescale which
in our subspace of $|\pm z\rangle=|\pm r\exp(i\psi)\rangle$ is given by
$1/r^2$. For initial $r\gg 1$ this decoherence is much faster than damping
and it takes place much before the states $|\pm z\rangle$ decay to the
ground state. In the opposite case of $r\ll 1$ the states $|\pm z\rangle$
decay to the ground state before they can be distinguished by the
environment.  Both limits were considered in Ref. \cite{exact}. Coherent
states with coherent amplitudes $\pm z$ that differ less than $1$ cannot
be distinguished. Here we concentrate on the distiguishable case of $r\gg
1$. In this limit we can self-consistently ignore damping and focus on the
decoherence and measurement process. In the homodyne limit ($R\gg r$),
where detection rates are fast as compared to the spontaneous emission
(decoherence) time, at any given time the correlators for the increments
$dn_i \equiv dN_i-\overline{dN_i(\rho)}$ are

\begin{eqnarray}
&& \overline{ dn_i }=0                  \;\;,\nonumber\\
&& \overline{ dn_i dn_j }\approx
   \delta_{i j} \; \overline{ dN_i } \approx
   \delta_{i j} \; \eta_i R^2 dt + {\cal O}(R) \;.
\end{eqnarray}
In this limit the increments can be approximated by $dn_i=\sqrt{\eta_i}
R\; dW_i$, where $dW_i$'s are gaussian Wiener increments, such that
$\overline{dW_i}=0$ and $\overline{dW_i dW_j}=\delta_{i j}dt$
\cite{wiseman,wisemanPhD}.  After introducing a variable $B$ as $A=\tanh
B$, and translating to Stratonovich convention, the super-observer's
equation for $B$ reads

\begin{equation}\label{dB}
\frac{dB}{d\tau} = \eta \tanh B + 
                   \sum_i \sqrt{\eta_i} \theta_i(\tau) ,
\label{dbdtau}
\end{equation}
where we have defined a new time variable $\tau = 4 t
r^2\cos^2(\phi-\psi)$
and $\eta=\sum_i \eta_i$. Here $\theta_i$ are 
stochastic continuous
functions of time, defined as $dW_i=\theta_i dt$. These
stochastic functions are white noises with correlators

\begin{equation}
\overline{ \theta_i(\tau_1) \;
           \theta_j(\tau_2)}  = \delta_{i j} \; 
\delta(\tau_1-\tau_2). 
\end{equation}
According to Eq. (\ref{dB}), $B$ initially performs a random walk driven
by the noises but once it diffuses into a positive ($\tanh B=+1$) or
negative ($\tanh B=-1$) domain, the deterministic force $\;\eta\tanh B\;$
takes over and inevitably drives $B$ towards positive or negative
infinity. After the transient time $\tau \sim 1/\eta$, $A$ settles down at
$A=\pm 1$ which corresponds to the pure state $|\pm z\rangle$ (see the
super-observer's trajectory $A(\tau)$ in Fig.1). By this time an observer
who knows all $dN_i(t)$ can tell whether the
system oscillator is in the state $|z\rangle$ or $|-z\rangle$, and attributes
to the system the appropriate pure state. This happens also 
when the total efficiency $\eta$ is less than one.
  
An observer $i$ ascribes to the system a state $\rho_i$ conditioned on his
own records $dN_i(t)$ only. Since we want to study correlations between
the measurement records, the evolution of $\rho_i$ is given by the MCSME
(Eq. (\ref{QBM_SOCME})). Taking the homodyne $R\gg r$ limit in the single
observer case we get

\begin{equation}
dN_i-\overline{dN_i(\rho_i)}=  
\left( dN_i-\overline{dN_i(\rho)} \right)+
\left( \overline{dN_i(\rho)}
       -\overline{dN_i(\rho_i)} \right)
\stackrel{R\gg r}{\approx}
\sqrt{\eta_i} R\; dW_i+
 2\eta_i r R\; (A-A_i)\cos(\psi-\phi).
\end{equation}
Just as for the case of the super-observer (Eq. (\ref{QBM_MOCME})), 
substitution of the
ansatz Eq. (\ref{AC}) into Eq. (\ref{QBM_SOCME}) and neglecting any
${\cal O}(1/R)$ terms, leads to the equation for the single observer
$B_i$,

\begin{equation}\label{dBS}
\frac{dB_i}{d\tau} = 
\eta_i \tanh B_i +
\left[ 
\eta_i \tanh B - \eta_i \tanh B_i+  
\sqrt{\eta_i} \; \theta_i
\right]=
\eta_i \tanh B +
\sqrt{\eta_i} \; \theta_i ,
\end{equation}
where $A_i=\tanh B_i$. The terms in the square brackets come from the
stochastic term in Eq. (\ref{QBM_SOCME}). Note that the super-observer's
$B$ appears in the evolution equation of the $B_i$ associated to the
single measurement channel $i$. This reflects the fact that the single
channel and multiple channels measurement results are correlated in the
MCSME (Eq. (\ref{MCSME})).

Let us now study how the evolutions of $A$ according to the super-observer
and single observers are correlated. On the one hand, according to Eq.
(\ref{dbdtau}), the super-observer evolution settles $A=\tanh B$ at $\pm
1$ after the transient time $\tau \sim 1/\eta$. On the other hand, the
single observer evolution is given by Eq. (\ref{dBS}), and correlations
between the two evolutions enter through the first term in the most right
hand side of that equation containing the super-observer $A=\tanh B$. Once
$A=\tanh B=\pm 1$ is chosen, the deterministic drift term $\eta_i \tanh
B\;$ on the RHS of Eq. (\ref{dBS}) will inevitably force $A_i=\tanh B_i$
to make the same choice after the longer transient time $\tau\sim
1/\eta_i$. Eventually all observers will settle down at $A=A_i=\pm 1$, and
the average relative purities will be equal to one, $O_i = O_{i j}=1$ (see
the single realizations for two measurement channels $i=1,2$ in Fig.1).

\begin{figure}
\centering \leavevmode \epsfxsize=7cm
\epsfbox{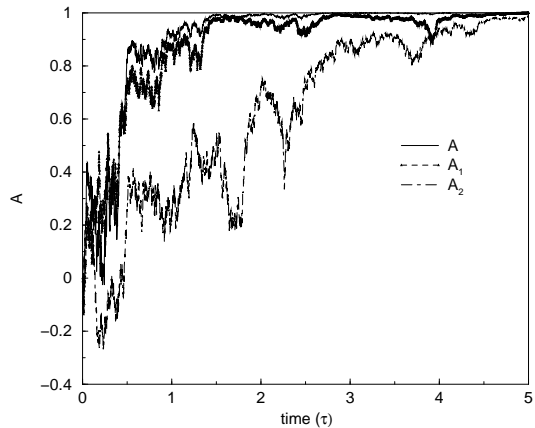} 
\caption{ A single realization of the stochastic trajectories $A(\tau)$ 
(thick line), $A_1(\tau)$ and $A_2(\tau)$ (thin lines) 
for $\eta_1=0.7$ and $\eta_2=0.3$. The super-observer's $A(\tau)$ settles 
at $+1$ around $\tau\approx 1$, it is
followed by $A_1(\tau)$ after a $\tau$-delay $\approx1$. $A_2(\tau)$ after
a long period of indecision settles down at $+1$ at $\tau \approx 5$. }
\end{figure}

In our example the observers finally find out which of the two coherent 
states is the state of the system. It is possible because the initial
coherent states have large amplitudes $\pm z$ with $|z|\gg 1$ so that
the decoherence time is much shorter than the spontaneous emission time.
In the opposite regime of $|z|\ll 1$ the decoherence time is longer than
the spontaneous emission time, and the observers will not find out the 
state before it decays to the vacuum state.

\subsection{Independence of the outcome distributions}

  To have a better feeling of the multiple channels stochastic master
equation (MCSME) formalism we consider the following example. There are
two observers $i=1,2$.  The measurements by observer $2$ affect the
evolution of the super-observer's density matrix $\rho$.  Since the
environment is also monitored by observer $1$, in principle observer $1$
may be able to identify perturbations of $\rho$ produced by measurements
of observer $2$ and realize that there is another observer monitoring the
system. In Section II we gave a general argument that, as a direct
consequence of the quantum Markov approximation, observer 2 cannot find
out if there is another observer. Here we present a simple calculation
which illustrates this fact in our example of superposition of coherent
states.

  To begin with, note that Eq. (\ref{dB}) is equivalent to the following
Fokker-Planck equation for the probability distribution $P(\tau,B)$ for
$B$ at time $\tau$ \cite{vankampen},

\begin{equation}\label{FP} 
\frac{1}{\eta}\frac{\partial P}{\partial\tau}=
-\frac{\partial}{\partial B} \tanh(B) P +
\frac{1}{2} \frac{\partial^2}{\partial B^2} P \;\;. 
\end{equation} 
We can compare the following two situations:

1) Observer $1$ is the only observer or $\eta_2=0$. His probability
distribution evolves according to Eq. (\ref{FP}). The initial condition
$P(0,B_1)=\delta(B_1)$ leads to the solution

\begin{equation}\label{Proba}
P^{(1)}(\tau,B_1)=
\frac{\exp\left( -\frac{(B_1-\eta_1\tau)^2}
                       {2\eta_1\tau}       \right)}
     {\sqrt{2\pi\eta_1\tau}}+
\frac{\exp\left( -\frac{(B_1+\eta_1\tau)^2}
                       {2\eta_1\tau}       \right)}
     {\sqrt{2\pi\eta_1\tau}} \;.
\end{equation}

2) There is an observer $2$ with $\eta_2 \gg \eta_1$. In this limit,
where the perturbations by observer $2$ are the strongest, one is most 
likely to suspect that the less efficient observer $1$ could find out about
the more efficient observer $2$. The evolution of $B(\tau)$ is 
mainly conditioned upon the measurements of observer $2$. The multiple
observer $A$ 
settles at $\pm 1$ on a time-scale $1/\eta$ which is much faster than the 
time $1/\eta_1$ observer $1$ needs to find out about the system. 
The state of the system is settled without any influence of measurements of 
observer $1$. Suppose that, with the probability $1/2$, the multiple
observer state 
$\tanh B=+1$ was chosen. $B_1$ evolves according to Eq. (\ref{dBS}) with a 
fixed $\tanh B=+1$. The probability distribution for $B_1$ is

\begin{equation}
P_+(\tau,B_1)=
\frac{\exp\left( -\frac{(B_1-\eta_1\tau)^2}
                       {2\eta_1\tau}       \right)}
     {\sqrt{2\pi\eta_1\tau}} \;.
\end{equation}
Also with the probability $1/2$, $\tanh B=-1$ can be chosen. Now $\tanh
B=-1$ is fixed in Eq. (\ref{dBS}) and the probability distribution is

\begin{equation}
P_-(\tau,B_1)=
\frac{\exp\left( -\frac{(B_1+\eta_1\tau)^2}
                       {2\eta_1\tau}       \right)}
     {\sqrt{2\pi\eta_1\tau}} \;.
\end{equation}
As we do not know which super-observer's state will be chosen, the two
probabilities add to give $P^{(2)}=P_+ + P_-$. It is easy to check that
$P^{(2)}=P^{(1)}$ in Eq. (\ref{Proba}). The probability distributions
coincide, so observer $1$ cannot find out if there is any observer $2$
even if observer 1 detects just $1\%$ of phonons and the other more
efficient observer detects $99\%$ or almost all phonons.


\section{ Two-level atom: Direct photo-detection }

We want to contrast the quantum Brownian motion model with an example of a
system with a small Hilbert space, such as a driven two-level atom coupled 
to the radiation field, for which we do not expect perfect pointer states. 
In Appendix A we derive the MCSME for a two-level atom driven
by a laser beam with frequency $\omega$ and whose emitted radiation is 
subjected 
to photo-detection. It takes the form

\begin{eqnarray}\label{2LACMEO}
&&
d\rho=
   - i \; dt \; \left[ \omega\sigma_x , \rho \right] +
   dt \; \left( c \rho c^{\dagger} -
                \frac{1}{2} c^{\dagger}c \rho -
                \frac{1}{2} \rho c^{\dagger}c  \right) + 
\sum_i \left( dN_i-\overline{dN_i(\rho)} \right) \;
\left( \frac{           c \rho c^{\dagger}    }
            { {\rm Tr}[ c \rho c^{\dagger} ]  }
       -\rho
\right)\;,
\label{2LA_MOCME}\\
&&
d\rho_i=
   - i \; dt \; \left[ \omega\sigma_x , \rho_i \right] +
   dt \; \left( c \rho_i c^{\dagger} -
                \frac{1}{2} c^{\dagger}c \rho_i -
                \frac{1}{2} \rho_i c^{\dagger}c  \right) +
\left( dN_i-\overline{dN_i(\rho_i)} \right)\;
\left( \frac{           c \rho_i c^{\dagger}    }
            { {\rm Tr}[ c \rho_i c^{\dagger} ]  }
       -\rho_i
\right)\; ,
\label{2LA_SOCME} \\
&&
\overline{dN_i(\rho)} =
\eta_i dt
{\rm Tr}[\rho c^{\dagger}c]\; .
\label{2LA_dN}
\end{eqnarray}
The density matrix $\rho$ of the atom is a $2\times 2$ matrix

\begin{equation}\label{rho}
\rho \;=\; \frac{1}{2}[I+x\sigma_x+y\sigma_y+z\sigma_z]\;.
\end{equation}
The lowering operator is $c=(\sigma_x-i\sigma_y)/2$, and 
the number of photons detected in channel $i$ between $t$ and $t+dt$ is 
$dN_i\in \{0,1\}$ with an average proportional to the occupation number of the
atom, see Eq. (\ref{2LA_dN}), and $\overline{dN_i dN_j}=
\delta_{i j}\overline{dN_i}$. Following each detection of a photon
(any $dN_i=1$), the atom is known to be in the ground
state (the $-1$ eigenstate of $\sigma_z$), from where it is excited again
by a laser beam through the Hamiltonian term $\omega\sigma_x$. The
efficiency $\eta_i$ of the detector used by observer $i$ is
the fraction of photons which are detected by him.

  When $\omega\gg 1$ the most predictable states of the two-level atom are $\sigma_x$
eigenstates, i.e., they are determined by the Hamiltonian $\omega\sigma_x$
describing the excitations via the laser beam \cite{zoe}. These 
states are far from perfect since they have a nonzero initial rate of purity
loss. Moreover, while eigenstates of $\sigma_x$ are most predictable, they are not
the most effective in making an imprint on the environment (as real pointer states
should be \cite{whz01,Zurek81,pointer}. In particular, the environment-system
Hamiltonian does not preserve them.
As a consequence, we do not expect agreement between observers
even if they are measuring in a basis of the environment correlated to the
$\sigma_x$-eigenstates of the atom. Direct photo-detection is a way
to find out if the atom is in the ground state. This state is complementary 
to the most predictable states. That is why we expect
the relative purity between observers to be very poor. In fact we 
will find any two observers to be anti-correlated, $O_{i j}<1/2$.


\subsection{The  $\omega\gg 1$ limit }

For $\omega\gg 1$ Eqs.(\ref{2LA_MOCME},\ref{2LA_SOCME}) can be solved
rigorously. Suppose that no photons are detected for a certain period of
time, $dN_{i}(t)=0$. During this time the density matrix $\rho$ in
Eq. (\ref{rho}) evolves according to the deterministic part of 
Eq. (\ref{2LA_MOCME}). The unitary self-evolution with the Hamiltonian
$\omega\sigma_x$ is mixing $y$ and $z$ with the frequency $2\omega$. It is
convenient to use the interaction picture, where

\begin{eqnarray}
x &=& x_{\rm int} , \nonumber \\
y &=& y_{\rm int}\cos 2\omega t-z_{\rm int}\sin 2\omega t , 
\label{interaction}\\
z &=& y_{\rm int}\sin 2\omega t+z_{\rm int}\cos 2\omega t , \nonumber
\end{eqnarray}
and the variation in time of $x_{\rm int},y_{\rm int},z_{\rm int}$ is slow as
compared to $\omega$. When we substitute the density matrix Eq. (\ref{rho}) into
the deterministic part of Eq. (\ref{2LA_MOCME}), use the interaction picture,
and average over one period of oscillation with frequency $\omega$, we
will obtain the following equations
\footnote{In the interaction picture the nonlinear terms in 
Eq. (\ref{2LA_MOCME}) average to zero thanks to the fast oscillations
with frequency $\omega$.}

\begin{eqnarray}
\label{drhoint}
\frac{dx_{\rm int}}{dt} &=& -\frac{1}{2} x_{\rm int} , \nonumber  \\
\frac{dy_{\rm int}}{dt} &=& -\frac{3}{4}(1-\eta) y_{\rm int} , \\
\frac{dz_{\rm int}}{dt} &=& -\frac{3}{4}(1-\eta) z_{\rm int} . \nonumber  
\end{eqnarray}
Every time a photon is detected the super-observer density matrix $\rho$ is 
projected to the ground state. All the information about the previous 
evolution of $\rho(t)$ is forgotten. Suppose that a detection took place at
time $t=0$. Just after the detection the initial conditions are
$x(0^+)=0,y(0^+)=0$ and $z(0^+)=-1$. Before the next detection happens, $x,y,z$
evolve according to Eqs.(\ref{interaction},\ref{drhoint}),

\begin{eqnarray}
\label{XYZ}
X(t) &=&0 , \nonumber \\
Y(t) &=& e^{-\frac{3}{4}(1-\eta)t}\sin 2\omega t ,  \\
Z(t) &=& -e^{-\frac{3}{4}(1-\eta)t}\cos 2\omega t , \nonumber 
\end{eqnarray}
where $t$ is the time elapsed since the last photo-detection. This solution
is valid until the next detection takes place. The next detection at $t=t_d$
will bring $\rho$ to the ground state again, from where the system will be
excited according to $x=X(t-t_d),y=Y(t-t_d),z=Z(t-t_d)$. The probability
that an observer $i$ will detect a photon between $t$ and $t+dt$
after the last detection by any observer is

\begin{equation}
\label{rate}
\overline{dN_i(\rho(t))} \;=\;
dt\;\eta_i \;\frac{1+Z(t)}{2}\;.
\end{equation}
The above argument can also be applied to Eq. (\ref{2LA_SOCME}).
Every time an observer $i$ detects a photon his state $\rho_i$
jumps to the ground state, from where it is excited according to

\begin{eqnarray}
\label{XYZalpha}
X_i(t) &=&0 , \nonumber   \\
Y_i(t) &=& e^{-\frac{3}{4}(1-\eta_i)t}\sin 2\omega t ,  \\
Z_i(t) &=& -e^{-\frac{3}{4}(1-\eta_i)t}\cos 2\omega t .
\nonumber
\end{eqnarray}
The time $t$ here is the time since the last detection by the observer $i$.


\subsection{ Distribution of waiting times }

In this example we shall see again  that a single observer
cannot find out if there is any other observer. We will consider just  
two observers $i=1,2$ and we will derive the distribution of waiting
times (times between subsequent detections) for observer $1$. We will show
that this distribution does not depend on $\eta_2$ so it is not sensitive
to the presence or absence of any observer 2. Any higher order
correlations between detection times can be expressed by this distribution
of waiting times because every time a photon is detected by observer $1$,
the atomic state goes down to the ground state so that any history before
the detection does not affect evolution that follows that detection. The
distribution of waiting times contains all the information observer 1 can
possibly extract from his measurements.

  Suppose that observer $1$ detects a photon at time $t=0$. What is the
probability $w_1(\tau)$ that he will detect the next photon at time
$t=\tau$ ? If observer $1$ were the only observer, so that $\eta_1=\eta$,
then the answer would be

\begin{equation}\label{palpha}
w_1(\tau)=
   \left( \eta_1\frac{1+Z_1(\tau)}{2} \right)
   e^{-\int_0^{\tau} d\tau_1\;
          \eta_1\frac{1+Z_1(\tau_1)}{2}}
\stackrel{\omega\gg 1}{\approx}
   \frac{\eta_1}{2}
   e^{-\frac{\eta_1\tau}{2}}.
\end{equation} 
The first factor is the average detection rate Eq. (\ref{rate}), and the
second one is the probability that no photon is detected between $0$ and
$\tau$. As it should be, $w_1(\tau)$ is normalized to unity. 
To obtain the final expression
for $w_1(\tau)$ in Eq. (\ref{palpha}) we have neglected all terms which 
vanish for $\omega\gg 1$ as well as fast oscillating terms 
$\sim \cos 2\omega\tau$.

If there is a second observer, then the detection rate of observer $i$ 
depends not on $z_{\alpha}(t)$ but on $z(t)$ (see Eq. (\ref{rate})). 
In general there may be $n=0,1,\dots,\infty$ detections by
observer $2$ between $0$ and $\tau$. Every time there is a
detection by observer $2$ at $t=t_j$, $(j=1,\dots,n)$, $z(t)$ jumps
down to $-1$. For $t_j<t<t_{j+1}$ it evolves as $z(t)=Z(t-t_j)$. The
probability that there is no detection by observer $1$ between times $0$
and $\tau$, given that there are $n$ detections by observer $2$ at the
times $t_1,\dots,t_n$, is given by

\begin{equation}\label{wt} 
D_n(t_1,\dots,t_n,\tau)=   
q_2(t_1)q_2(t_2-t_1)\dots q_2(t_n-t_{n-1}) 
e^{-\frac{\eta}{2}(\tau-t_n)},
\end{equation} 
where $q_2(\tau)$ is distribution of waiting times for observer $2$ given
that there are no detections by $1$,

\begin{equation}\label{q2}
q_2(\tau)
\stackrel{\omega\gg 1}{\approx}
  e^{-\frac{\eta}{2}\tau}
  \frac{\eta_2}{2}
  (1-e^{-\frac{3}{4}(1-\eta)\tau}\cos 2\omega \tau),
\end{equation}
and the factor $e^{-\frac{\eta}{2}(t-t_n)}$ is the probability that no
detections by any observer take place between $t_n$ and $t$. The
distribution of waiting times for observer $1$, averaged over
detections by observer $2$, is given by $D_n$ multiplied by the detection
rate of observer 1 at $\tau$, and averaged over all possible $n$ and
$t_1,\dots,t_n$. Therefore the final expression for the waiting time 
distribution $f_{{\rm wait}}(\tau)$ 
for observer 1 in the presence of detections by observer 2 is
given by

\begin{equation}\label{tildap}
f_{{\rm wait}}(\tau) = \sum_{n=0}^{\infty}
  \int_0^{\tau} dt_1 
  \int_{t_1}^{\tau} dt_2 \dots 
  \int_{t_{n-1}}^{\tau} dt_n\;
  D_n(t_1,\dots,t_n,\tau)
  \frac{\eta_1}{2}[ 1 + Z(\tau-t_n) ]
\stackrel{\omega\gg 1}{\approx}
\frac{\eta_1}{2}e^{ -\frac{\eta_1}{2}\tau},
\end{equation}
where, again, we have neglected terms which vanish for $\omega\gg 1$ and
any fast oscillating terms. In Appendix B we show how to obtain this last
formula. We conclude that the distribution of
waiting times for observer $1$ in the presence of detections by observer
2 (Eq. (\ref{tildap})) is the same as that for no observer 2 present 
(Eq. (\ref{palpha})). The distribution of waiting times for observer 1 is
not sensitive to observer $2$.


\subsection{Average relative purity between $\rho_1$ and $\rho$ }

Let us now find out how much does a given observer, say $i=1$, know
about the state of the super-observer. To this end we will calculate the
average relative purity between the single observer and the super-observer 
density
matrices, $O_1=\overline{{\rm Tr} \rho_1 \rho}$. Imagine the
following situation. Take an arbitrary instant of time $\tau=0$ and 
call $o_1^{(n)}(\tau=0)={\rm Tr} \rho_1(\tau=0) \rho(\tau=0)$ the 
relative purity given that the last detection of observer 1 took place
at $\tau= -t$ and there were $n$ detections by observer $2$ between $\tau=-t$
and $\tau=0$. Then the average relative purity
$O_1$ evaluated at the time $\tau=0$ will be equal to the $t$-average 
(i.e., average over all possible initial times of detection by 1) of the 
relative purity $o_1^{(n)}(\tau=0)$ 
given that there were no detections by observer $1$ between $\tau=-t$ and
$\tau=0$ and averaged over all the possible numbers $n$ of detections by
observer $2$ and his detection times $t_1,\dots,t_n$. For the sake of clarity,
we now shift the time origin as $\tau \rightarrow \tau + t$, so that the last
detection of 1 took place at time $0$ and we are interested in evaluating
$O_1$ at time $t$. The unnormalized probability distribution for no
detections by observer $1$ between $0$ and $t$, and $n$ detections by
observer $2$ at the times $t_1,\dots,t_n$ is $D_n(t_1,\dots,t_n,t)$, given
by Eq. (\ref{wt}). The normalizing factor for this distribution is

\begin{equation}
n_1=
\sum_{n=0}^{\infty}
  \int_0^{\infty} dt 
  \int_0^t dt_1 
  \int_{t_1}^t dt_2 \dots 
  \int_{t_{n-1}}^t dt_n\;
  D_n(t_1,\dots,t_n,t)
\stackrel{\omega\gg 1}{\approx}
  \frac{2}{\eta_1} \;.
\end{equation}
Given that the last detection by observer $1$ took place at time $0$
and the last detection by any observer happened at time $t_n$, the
relevant relative purity is

\begin{equation}
o_1^{(n)}(t)={\rm Tr} \rho_1(t)\rho(t)=
\frac12+\frac12 e^{-\frac34(1-\eta_1)t}
                e^{-\frac34(1-\eta)(t-t_n)}
                \cos 2\omega t_n \;.
\end{equation}
This relative purity, 
when averaged with the probability distribution (\ref{wt}),
gives

\begin{equation}
\label{ouno}
O_1=
n_1^{-1}
\sum_{n=0}^{\infty}
  \int_0^{\infty} dt
  \int_0^t dt_1
  \int_{t_1}^t dt_2 \dots
  \int_{t_{n-1}}^t dt_n\;
  D_n(t_1,\dots,t_n,t) \; o_1^{(n)}(t) = 
\frac{1}{2} + \frac{\eta_1}{2 (3-2\eta_1)} ,
\end{equation}
where we have neglected all ${\cal O}(1/\omega)$ terms.

On the other hand, the average purity gained by observer $1$ can be 
calculated as follows. According to Eqs.(\ref{XYZalpha}), the purity 
at the time $t$ after the last detection is 
$o_{11}(t)={\rm Tr} \rho_1^2(t)= \frac{1}{2}+\frac{1}{2}
\exp[-\frac{3}{2}(1-\eta_1)t]$. The probability that 
there was no detection between
$0$ and $t$ is $\exp(-\eta_1t/2)$. Therefore, the average 
purity $O_{11}$ is $o_{11}(t)$ averaged over $t$:

\begin{equation}
\label{puno}
O_{11}=\frac{ \int_0^{\infty}dt\; e^{-\frac{\eta_1}{2}t}\;o_{11}(t) }
            { \int_0^{\infty}dt\; e^{-\frac{\eta_1}{2}t}         }
=\frac{1}{2} + \frac{\eta_1}{2(3-2\eta_1)} \;.
\end{equation}
As expected from Eq. (\ref{Oii}), $O_{11}$ coincides with $O_1$ (see Fig.2). 
Let us now
comment on the limiting cases $\eta_1=0$ and $\eta_1=1$. In the former case
we get $O_{11}=0.5$, that corresponds to no information gain by the observer
($\rho_1$ is maximally mixed). In the latter case we get $O_{11}=1$, that
is maximal gain of information, and $\rho_1$ is pure.

\begin{figure}
\centering \leavevmode \epsfxsize=7cm
\epsfbox{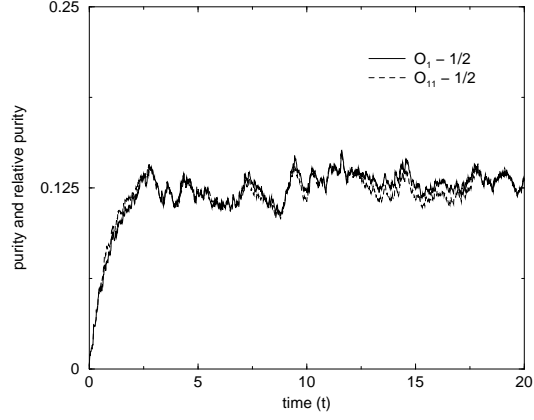} 
\caption{Average purity $O_{11} - 1/2$ and average relative purity
$O_1 - 1/2$ for an observer performing photo-detection measurements. 
The initial condition is maximal lack of knowledge, i.e. 
$\rho(t=0)=\rho_1(t=0)=I/2$. The super-observer's efficiency is 
$\eta=0.6$ and 
the single observer one is $\eta_1=0.5$. According to Eqs.(\ref{ouno}, \ref{puno}) 
the asymptotic value is $O_1(\infty)=O_{11}(\infty)= 0.125$. The stochastic trajectories 
are an average over 256 single realizations. The small discrepancy between $O_1$
and $O_{11}$ in the figure is an artifact of the finite number of realizations
used for calculating the averages. }
\end{figure}


\subsection{Average relative purity $O_{i j}$}

The average relative purity $O_{12}=\overline{ {\rm Tr} \rho_1 \rho_2}$ 
has contributions from the following two situations:

(1) The last detection by observer $1$ took place at time $0$. Between
times $0$ and $t$ there were $n \geq 1$ detections by observer $2$ at the
times $t_1,\dots,t_n$. The last detection before $t$ was made by observer
$2$ at time $t_n$. According to Eqs.(\ref{XYZ}), the relative purity at $t$ is

\begin{equation}
o_{12}^{(n)}(t)={\rm Tr} \rho_1(t)\rho_2(t)=
\frac12+\frac12 e^{-\frac34(1-\eta_1)t} e^{-\frac34(1-\eta_2)(t-t_n)} 
\cos 2\omega t_n \;.
\end{equation}
The normalizing factor for the probability distribution is

\begin{equation}
n_{12}=   
\sum_{n=1}^{\infty}
  \int_0^{\infty} dt
  \int_0^t dt_1
  \int_{t_1}^t dt_2 \dots
  \int_{t_{n-1}}^t dt_n\;
  D_n(t_1,\dots,t_n,t)
\stackrel{\omega\gg 1}{\approx}
  \frac{2 \eta_2}{\eta_1 (\eta_1 + \eta_2)} \;,
\end{equation}
and the averaged relative purity is

\begin{equation}
\overline{o}^{(1)}_{12}=
n_{12}^{-1}
\sum_{n=1}^{\infty}
  \int_0^{\infty} dt
  \int_0^t dt_1
  \int_{t_1}^t dt_2 \dots
  \int_{t_{n-1}}^t dt_n\;
  D_n(t_1,\dots,t_n,t) o_{12}^{(n)}(t) = 
\frac{1}{2} - \frac{\eta_1 (\eta_1+\eta_2)}{\eta_2 (6-\eta_1-\eta_2)
(7-\eta_2 - 4 \eta_1)} .
\end{equation}

(2) The last detection before $t$ was made by observer 1 instead of observer
2, as in the case (1). The description of this second situation is the same as 
above, except that observers $1$ and $2$ are interchanged. In particular,
the final result for the relative purity reads

\begin{equation}
\overline{o}^{(2)}_{12}=
\frac{1}{2} - \frac{\eta_2 (\eta_1+\eta_2)}{\eta_1 (6-\eta_1-\eta_2)
(7-\eta_1 - 4 \eta_2)} .
\end{equation}
In general $\eta_1\neq\eta_2$ and the two situations are not equally
likely. Let us call $p^{(1)}$ the probability that case (1) happens;
clearly for case (2) we have $p^{(2)}=1-p^{(1)}$. The probability $p^{(1)}$
is given by $p^{(1)}=n_{12} / n_1 = \eta_2/(\eta_1+\eta_2)$.
The relative purity averaged over the two situations is then

\begin{eqnarray}
\label{ounodos}
O_{12} &=& \frac{\eta_2}{\eta_1+\eta_2} \; \overline{o}^{(1)}_{12}+
         \frac{\eta_1}{\eta_1+\eta_2} \; \overline{o}^{(2)}_{12} \nonumber
\\
&=&
\frac{1}{2} - \frac{\eta_1 \eta_2 [ 6 -2(\eta_1+\eta_2)]}
{2 (6-\eta_1-\eta_2) (3-2 \eta_1) (3 -2 \eta_2)} .
\end{eqnarray}
In figures 3 and 4 we show simulations of the time evolution of the 
relative purity 
$O_{12}$ for the case $\eta_1=\eta_2$ and $\eta_1 \neq \eta_2$.

\begin{figure}
\centering \leavevmode \epsfxsize=7cm
\epsfbox{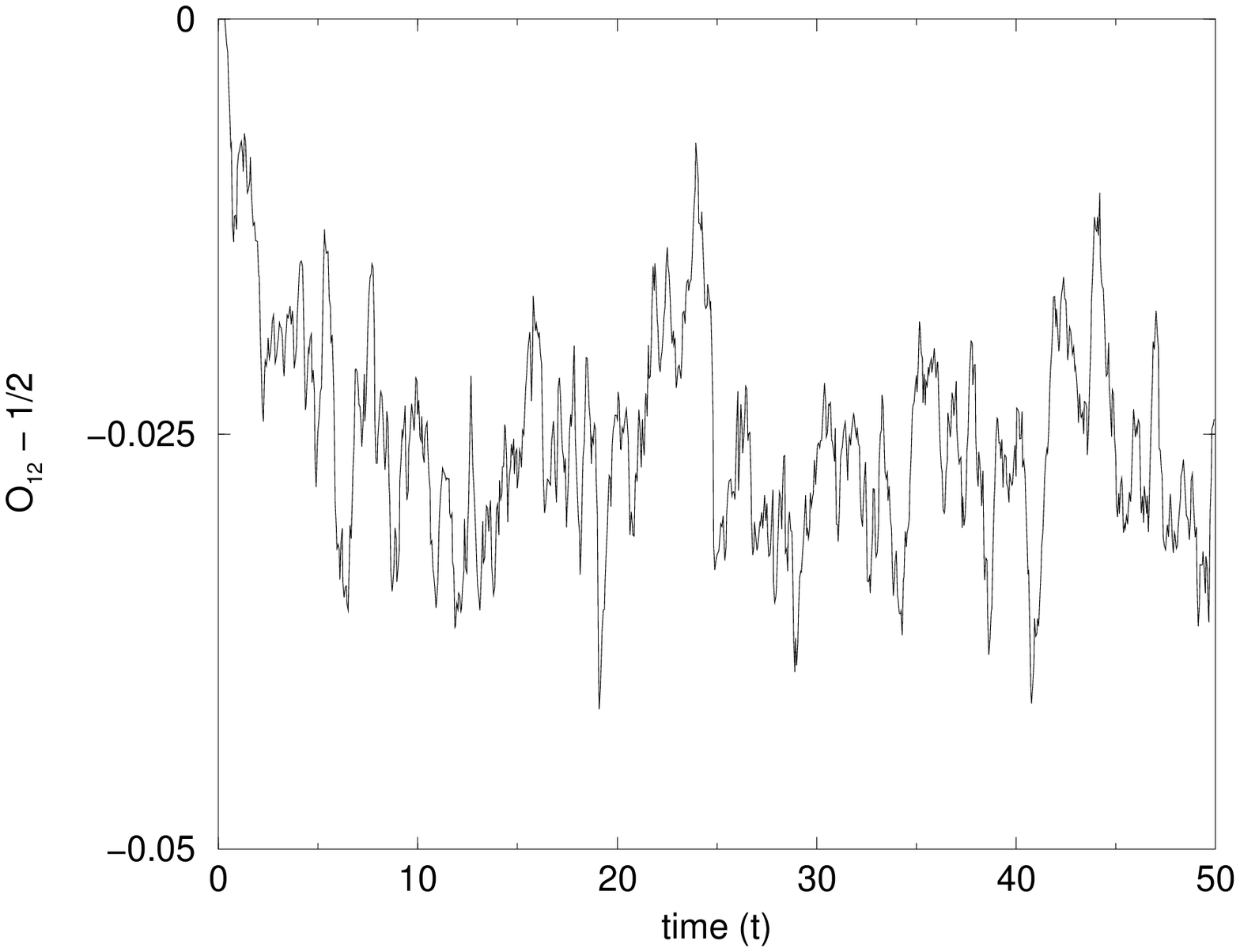} 
\caption{Average relative purity $O_{12} - 1/2$ between two observers 
performing photo-detection measurements. 
Their initial condition is maximal lack of knowledge, i.e. 
$\rho_1(t=0)=\rho_1(t=0)=I/2$. The efficiencies are $\eta_1=\eta_2=0.5$.
According to Eq. (\ref{ounodos}) the asymptotic value of the relative
purity 
is -0.025. The stochastic trajectory is an 
average over 256 single realizations. }
\end{figure}

\begin{figure}
\centering \leavevmode \epsfxsize=7cm
\epsfbox{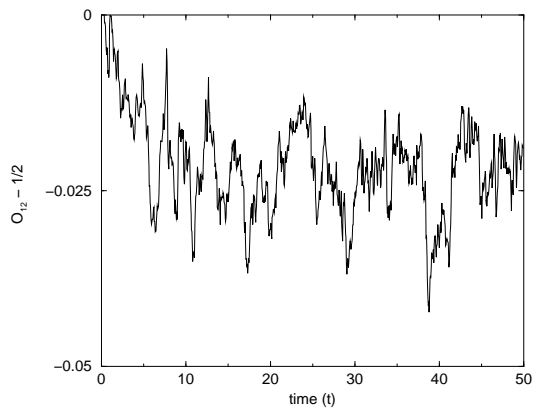} 
\caption{Average relative purity $O_{12} - 1/2$ between two observers 
performing photo-detection measurements. 
Their initial condition is maximal lack of knowledge, i.e. 
$\rho_1(t=0)=\rho_1(t=0)=I/2$. The efficiencies are $\eta_1=0.7$ and
$\eta_2=0.3$.
According to Eq. (\ref{ounodos}) the asymptotic value of the 
relative purity 
is -0.022.  The stochastic trajectory is an 
average over 256 single realizations.}
\end{figure}

Note that the average relative purity is manifestly less than $1/2$: the single
observer states
$\rho_1$ and $\rho_2$ are anti-correlated. The reason for this
anti-correlation can be explained as follows. Suppose that the states
$\rho_1, \rho_2, \rho$ are initially 
fully correlated (i.e., relative purity equal to one). 
Observer $1$ is most likely to have a 
detection when the super-observer's state is excited ($z\approx +1$). The 
hypothetical positive correlation means that when $z\approx +1$, then also 
$z_1\approx +1$ and $z_2\approx +1$. Suppose that a detection by observer 
$1$ happens. The super-observer $z$ and the single observer $z_1$ jump 
down to $-1$. 
The observer $2$ has no clue that there was a detection by observer $1$.  
What is more, the super-observer $z$ is close to $-1$ so observer $2$ cannot
detect a photon and jump to $z_2=-1$. His $z_2$ remains close to $+1$.
Just after the detection the product $\;z\;z_1\;>0$ but the product
$\;z_1\;z_2\;<0$. This mechanism cannot make $O_1<1/2$ but it can and it
does make $O_{12}< 1/2$.

  We have solved exactly the problem of correlations between multiple measurement
channels in the limit of $\omega\gg 1$. This limit is sufficient to illustrate our
ideas. However, the exact solution for arbitrary $\omega$ of the resonance
fluorescence problem in Ref.\cite{erfp} suggests that, with some extra
work, our formulas for average relative purities 
can be generalized exactly to arbitrary $\omega$.


\section{ Two-level atom: Homodyne detection }

As we saw in the previous section, direct photo-detection is a way to find
out if the atom is in the ground state. One can also measure different
quadratures of the two-level atom by performing homodyne detection on the
radiation emitted from it \cite{qoptics}. In general, it is possible to 
measure the
expectation value of the operator $(x\cos\phi -y\sin\phi)$, where $\phi$ is
the phase of the local oscillator in the homodyne detector. This kind of
measurement tends to localize the state of the atom around the eigenstates 
of the operator
$(\sigma_x\cos\phi-\sigma_y\sin\phi)$. The MCSME is (see Appendix A)

\begin{eqnarray}\label{HOMCMEO}
&&
d\rho = -i\;dt\left[ \omega\sigma_x , \rho \right] +
        dt \; \left( c \rho c^{\dagger} -
                     \frac{1}{2} c^{\dagger}c \rho -
                     \frac{1}{2} \rho c^{\dagger}c  \right) +
\sum_i
     \left( dN_i-\overline{dN_i(\rho)} \right) \;
     \left( \frac{   (c+\gamma_i)\rho
                     (c^{\dagger}+\gamma^{\star}_i)   }
     { {\rm Tr}[ (c+\gamma_i)\rho
                 (c^{\dagger}+\gamma^{\star}_i) ] }
                 -\rho  \right)\;,
\label{HOM_MOCME}\\
&&
d\rho_i = 
-i\;dt\left[ \omega\sigma_x , \rho_i \right] +
    dt \; \left( c \rho_i c^{\dagger} -
                 \frac{1}{2} c^{\dagger}c \rho_i -
                 \frac{1}{2} \rho_i c^{\dagger}c  \right) +
     \left( dN_i-\overline{dN_i(\rho_i)} \right) \;
     \left( \frac{   (c+\gamma_i)\rho_i
                     (c^{\dagger}+\gamma^{\star}_i)   }
     { {\rm Tr}[ (c+\gamma_i)\rho_i
                 (c^{\dagger}+\gamma^{\star}_i) ] }
                 -\rho_i  \right) \; ,
\label{HOM_SOCME} \\
&&
\overline{dN_i(\rho)} =
\eta_i dt [ R_i^2 + R_i e^{-i\phi_i} 
{\rm Tr}\rho c + R_i e^{+i\phi_i} {\rm Tr} \rho c^{\dagger}+
{\rm Tr} \rho c^{\dagger}c ] \;.
\label{HOM_dN}
\end{eqnarray}
Here $\gamma_i=R_i \exp(i\phi_i)$ is the complex
amplitude of the local oscillator of the detector $i$. We will
eventually take the limit $R_i \rightarrow\infty$. We allow each
observer to have his own homodyne phase $\phi_i$, so that they can
measure different quadratures, i.e., they measure non-commuting
observables (a related experimental realization of measurements of 
non-commuting observables in two channels in cavity QED was carried out
in \cite{foster}). The detector currents are proportional to
Eq. (\ref{HOM_dN}). The case $\phi_i=0$ corresponds to measurement of
the $x-$quadrature and $\phi_i=\pi/2$ to $y-$quadrature. The large
$R_i$ limit of Eq. (\ref{HOM_MOCME}) is

\begin{eqnarray}\label{dWCMEO}
d\rho &=& -i\;dt\left[ \omega\sigma_x , \rho \right] +
          dt \; \left( c \rho c^{\dagger} -
                       \frac{1}{2} c^{\dagger}c \rho -
                       \frac{1}{2} \rho c^{\dagger}c  \right) 
\nonumber\\
&& + \sum_i \sqrt{\eta_i} \;dW_i
\left[ c\rho e^{-i\phi_i} + \rho c^{\dagger}e^{+i\phi_i} -
\rho {\rm Tr} \left( c\rho e^{-i\phi_i} + 
\rho c^{\dagger}e^{+i\phi_i} \right) \right]\;,
\end{eqnarray}
where $dW_i$ are gaussian Wiener increments such that
$\overline{dW_i}=0$ and
$\overline{dW_i dW_j}=\delta_{i j}dt$.
To derive the large $R_i$ limit of Eq. (\ref{HOM_SOCME})
we first split $dN_i-\overline{dN_i(\rho_i)}= 
\left( dN_i - \overline{dN_i(\rho)} \right)+
\left( \overline{dN_i(\rho)} - \overline{dN_i(\rho_i)} \right)$.
In the large $R_i$ limit the first term is proportional to 
$R_i \sqrt{\eta_i} dW_i$, and the second term is 
proportional to 
$R_i \eta_i dt {\rm Tr} [c (\rho-\rho_i) e^{-i \phi_i} + 
(\rho-\rho_i) c^{\dagger} e^{+i \phi_i} ]$. The
large $R_i$ limit of Eq. (\ref{HOM_SOCME}) reads

\begin{eqnarray}\label{dWCMES}
d\rho_i &=& 
      -i\;dt\left[ \omega\sigma_x , \rho_i \right] +
          dt \; \left( c \rho_i c^{\dagger} -
                       \frac{1}{2} c^{\dagger}c \rho_i -   
                       \frac{1}{2} \rho_i c^{\dagger}c  \right) \\   
&&  + \left[ \sqrt{\eta_i} \; dW_i +
           \eta_i dt\;
           {\rm Tr}\;\left( c(\rho-\rho_i) e^{-i\phi_i} +
                      (\rho-\rho_i) c^{\dagger}e^{+i\phi_i}
               \right)
    \right] \;
    \left[ \left( c\rho_i e^{-i\phi_i} +
                  \rho_i c^{\dagger}e^{+i\phi_i} \right) -
           \rho
  {\rm Tr} \left( c\rho_i e^{-i\phi_i} +
  \rho_i c^{\dagger}e^{+i\phi_i} \right) \right] . \nonumber 
\end{eqnarray}


\subsection{Average relative purity between $\rho_i$ and $\rho$ }

Unfortunately it is not possible to find analytic solutions to the above
equations for all values of the efficiencies $\eta_i$. For small
values of these efficiencies it is possible to work out various 
relative purities by a perturbative expansion in powers of $\eta_i$. For 
$\eta_i=0$, the conditional master equation (Eq. (\ref{dWCMEO})) is the unconditional
master equation, which has a stationary solution $\rho_{ss}$. In the
limit $\omega \gg 1$ it is equal to $\rho_{ss}=I/2$ or 
$x_{ss}=y_{ss}=z_{ss}=0$. The full density matrix is perturbed from this
stationary state by the noises $d W_i$, and the magnitude of the
perturbation grows with $\eta_i$. We expand 
$\rho=\rho_{ss} + \delta \rho$, the last term containing those perturbations.
Let us write $\delta \rho= (x \sigma_x + y \sigma_y + z \sigma_z)/2$. 
We expand $x$ as $x=x^{(1)} + x^{(2)} + \ldots$, where $x^{(1)}$ is
of order $\eta_i^{1/2}$, $x^{(2)}$ is of order $\eta_i^{3/2}$, etc. Similar
expansions are used for $y$ and $z$. 
To first  order in $\eta_i^{1/2}$
Eq. (\ref{dWCMEO}) reads

\begin{eqnarray}
\frac{dx^{(1)}}{dt} &=&-\frac12 x^{(1)}
  +\sum_i \sqrt{\eta_i} \;
\theta_i \;\cos\phi_i ,
\nonumber \\
\frac{dy^{(1)}}{dt} &=& -\frac12 y^{(1)}-2\omega z^{(1)}
  -\sum_i \sqrt{\eta_i} \;
\theta_i \;\sin\phi_i ,\\
\frac{dz^{(1)}}{dt} &=& -z^{(1)}+2\omega y^{(1)} . \nonumber
\end{eqnarray}
These equations have a solution

\begin{eqnarray}
\label{solution}
x^{(1)}(t) &=& \sum_i x^{(1)}_i(t) , \nonumber \\
y^{(1)}(t) \pm iz^{(1)}(t) &=&
   \sum_i 
   \left[ y^{(1)}_i(t) \pm i z^{(1)}_i(t) \right] ,
\end{eqnarray}
where
\begin{eqnarray}
\label{solu}
x^{(1)}_i(t) &=& \sqrt{\eta_i}\;
\cos\phi_i \int_{-\infty}^t d\tau\; e^{-\frac12(t-\tau)}\;
\theta_i(\tau) , \nonumber \\
y^{(1)}_{i}(t) \pm i z^{(1)}_{i}(t) &=&
   - \sqrt{\eta_{i}} \;
   \sin\phi_{i}
   \int_{-\infty}^t d\tau\;
   e^{-(\frac{3}{4} \mp 2 i \omega)(t-\tau)}\;
   \theta_{i}(\tau) .
\end{eqnarray}
To leading order in $\eta_{i}^{1/2}$'s the single observer equation
(\ref{HOM_SOCME}) is

\begin{eqnarray}
\frac{dx^{(1)}_{i}}{dt} &=&
     -\frac{1}{2}x^{(1)}_{i}
     + \sqrt{\eta_{i}} \;
\theta_{i} \;\cos\phi_{i} , \nonumber \\
\frac{dy^{(1)}_{i}}{dt} &=&
     -\frac{1}{2}y^{(1)}_{i}
     -2\omega z^{(1)}_{i}
     - \sqrt{\eta_{i}} \;
\theta_{i} \;\sin\phi_{i} , \\
\frac{dz^{(1)}_{i}}{dt} &=&
     -z^{(1)}_{i}+2\omega y^{(1)}_{i} . \nonumber
\end{eqnarray}
These equations are solved by the already introduced
$x^{(1)}_{i},y^{(1)}_{i},z^{(1)}_{i}$. To leading order in
$\eta_{i}$'s the relative purity
$O_{i}=\overline{{\rm Tr} \rho_{i} \rho}$ is

\begin{equation}\label{Oalpha}
O_{i} \equiv
\frac12+
\frac12
\overline{ 
\left[
x^{(1)}x^{(1)}_{i} + y^{(1)}y^{(1)}_{i} + z^{(1)}z^{(1)}_{i}
\right] } =
\frac12+
\frac12
\overline{ 
\left[
x^{(1)}_{i}x^{(1)}_{i} + y^{(1)}_{i}y^{(1)}_{i} +    
z^{(1)}_{i}z^{(1)}_{i}  \right] } .
\end{equation}
A straightforward calculation leads to the following stationary 
average relative purity

\begin{equation}
\label{overhomo}
O_{i}=
\frac12+  
\eta_{i}\left[ \frac12 \cos^2\phi_{i} +
                    \frac13 \sin^2\phi_{i} \right] .
\end{equation}
As we can see from Eq. (\ref{Oalpha}) the average relative purity coincides with
the average purity $O_{ii}$. The latter is the highest for
measurement basis correlated to the pointer states basis of the system, 
i.e., when 
 $\phi_{i}=0$. Through this measurement one can find out most about 
the system. In figures 5 and 6 we plot the average relative purity $O_1$ and the 
average purity $O_{11}$ for different values of the efficiencies and 
homodyne phases.

\begin{figure}
\centering \leavevmode \epsfxsize=7cm
\epsfbox{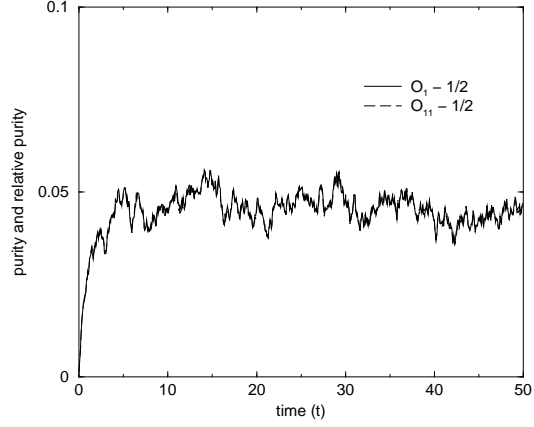} 
\caption{Average purity $O_{11} - 1/2$ and average relative purity
$O_{1} - 1/2$ for an observer
performing homodyne measurements. The efficiency is $\eta_1=0.1$ and
the homodyne phase is $\phi_1=0$.
According to Eq. (\ref{overhomo}), which is valid for small efficiencies,
the asymptotic value of the average relative purity and average purity is
$O_1-1/2 = O_{11}-1/2 =  0.05$. In the scale
of the figure $O_1$ and $O_{11}$ practically coincide. The stochastic 
trajectories are an average over 256 single realizations. }
\end{figure}

\begin{figure}
\centering \leavevmode \epsfxsize=7cm
\epsfbox{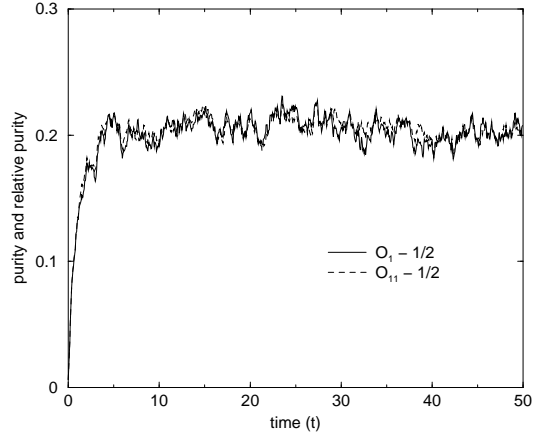} 
\caption{Average purity $O_{11}-1/2$ and average relative purity
$O_{1}-1/2$ for an observer
performing homodyne measurements. The efficiency is $\eta_1=0.5$ and
the homodyne phase is $\phi_1=0$. We do not expect Eq. (\ref{overhomo})
to hold for such a big efficiency. The stochastic trajectories are an 
average over 256 single realizations. The small discrepancy between $O_1$
and $O_{11}$ in the figure is an artifact of the finite number of trajectories used
to calculate the averages.}
\end{figure}


\subsection{Average relative purity $O_{ij}$}

The average relative purity 
$O_{12}=\overline{{\rm Tr}\rho_{1}\rho_{2}}$
is zero to leading order in $\eta_{i}$. To get a nonzero average relative
purity
we have to go one step further in the perturbative expansion for
$x$, $y$, and $z$. The equations for the second order terms that follow from
the single observer equation Eq. (\ref{dWCMES}) are

\begin{eqnarray}
\frac{dx^{(2)}_{i}}{dt} &=&
     -\frac12 x^{(2)}_{i}
     + \eta_{i}  \cos^2\phi_{i}   x^{(1)}_{j} 
     - \eta_{i}  \sin\phi_{i} \cos\phi_{i} y^{(1)}_{j}, 
     \nonumber \\
\frac{dy^{(2)}_{i}}{dt} &=&
     -\frac12 y^{(2)}_{i}
     -2\omega z^{(2)}_{i}
     + \eta_{i} \sin^2\phi_{i} y^{(1)}_{j} 
     - \eta_{i}  \sin\phi_{i} \cos\phi_{i} x^{(1)}_{j} ,
\\
\frac{dz^{(2)}_{i}}{dt} &=&
     -z^{(2)}_{i}+2\omega y^{(2)}_{i} . \nonumber
\end{eqnarray}
Formal solutions of these equations are

\begin{eqnarray}
\label{sol2level}
x^{(2)}_{i}(t) &=&
   \eta_{i} \int_{-\infty}^t d\tau\;
   e^{-\frac12(t-\tau)} 
        \left(
\cos^2\phi_{i}  x^{(1)}_{j}(\tau) - 
\sin\phi_{i} \cos\phi_{i} y^{(1)}_{j}(\tau) 
        \right) ,  \nonumber \\
y^{(2)}_{i}(t) \pm i z^{(2)}_{i}(t) &=&
   \eta_{i} \int_{-\infty}^t d\tau\;
   e^{-\frac34(t-\tau) \pm 2 i \omega (t-\tau)}\;  
        \left(
\sin^2\phi_{i} y^{(1)}_{j}(\tau) -
\sin\phi_{i} \cos\phi_{i} x^{(1)}_{j}(\tau) 
        \right).
\end{eqnarray}
To the first non-vanishing order in $\eta_{i}$ the average relative
purity is 

\begin{equation}
\label{o2level}
O_{12} =
\frac12 + 
\frac12 \overline{ 
\left[ 
x_{1}x_{2} + y_{1}y_{2} +
z_{1}z_{2} \right] }=
\frac12 +
\frac12 \overline{ 
\left[ 
x_{1}^{(1)}x_{2}^{(2)} +
                          y_{1}^{(1)}y_{2}^{(2)} +
                          z_{1}^{(1)}z_{2}^{(2)}  \right] }+
\frac12 \overline{ 
\left[
x_{1}^{(2)}x_{2}^{(1)} +
                          y_{1}^{(2)}y_{2}^{(1)} +
                          z_{1}^{(2)}z_{2}^{(1)} \right] }\;.
\end{equation}
We evaluate this expression in Appendix C. The result is

\begin{equation}
\label{over2level}
O_{12}=
\frac12+
\eta_{1}\eta_{2}
\left[ \cos^2\phi_{1}\cos^2\phi_{2}+
       \frac{4}{9} \sin^2\phi_{1}\sin^2\phi_{2} \right] .
\end{equation}
The average relative purity is maximized when both observers perform $x$-measurements
($\phi_{1}=\phi_{2}=0$). 
We verified this formula by numerical simulations using $\eta_1=\eta_2=0.01$.
Below, in figures 7 and 8, we plot the average relative purity 
$O_{12}$ for different sets of
homodyne phases and efficiencies $\eta_1=\eta_2=0.1$. These efficiencies are 
beyond the range of validity of Eq. (\ref{over2level}).

\begin{figure}
\centering \leavevmode \epsfxsize=7cm
\epsfbox{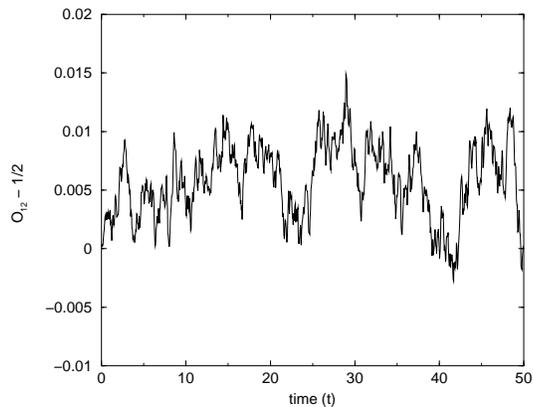} 
\caption{Average relative purity $O_{12} - 1/2$ between two observers
performing homodyne measurements. The efficiencies are $\eta_1=\eta_2=0.1$
and 
the homodyne phases are $\phi_1=\phi_2=0$. The stochastic trajectory is an 
average over 256 single realizations.}
\end{figure}

\begin{figure}
\centering \leavevmode \epsfxsize=7cm
\epsfbox{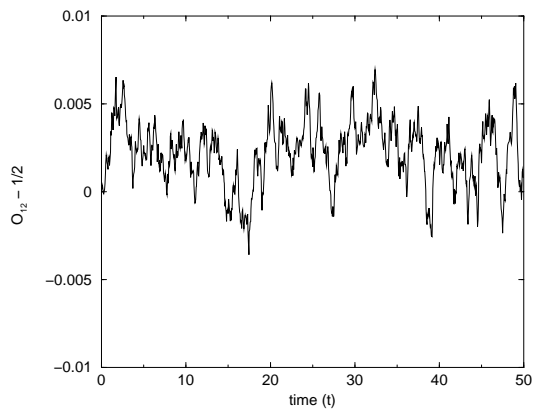} 
\caption{Average relative purity $O_{12} - 1/2$ between two observers
performing homodyne measurements. The efficiencies are $\eta_1=\eta_2=0.1$
and 
the homodyne phases are $\phi_1=\phi_2=\pi/2$. The stochastic trajectory is an 
average over 256 single realizations.}
\end{figure}


\section{ Concluding remarks }  

   Let us summarize the new results contained in this paper. We have
studied continuous quantum measurement with several observers and
we have demonstrated that it reduces to the ``single observer'' case.
The key problem of consistency of the sets of data acquired by different
observers is then reduced to the probability that a given combination
of data sets will be ever detected by the super-observer. We have applied the
formalism to several examples of quantum optics as well as to quantum
Brownian motion. Observers gain information about the state of the system
from their measurement records.  We have shown that observers gain most
information about the system and they agree the most when they measure in
environmental basis most correlated to the pointer basis of the system.  

Several questions regarding correlations between measurement records of
different observers were posed. We have shown that the problem of consistency
of sets of data acquired by different observers is reduced to the probability
that a given combination of data sets will ever be detected by the super-observer.
We have introduced average relative purity to study correlations between measurement
records of different observers. For the model of zero temperature quantum Brownian
motion (which is equivalent to the model of a damped harmonic oscillator) coherent
states are perfect pointer states: The solution to Eq. (\ref{QBMCMEO}) for an initial
coherent state remains pure and it is just a coherent state with decaying amplitude. We 
have shown that for an initial Schr\"odinger cat state $(|z\rangle+|-z\rangle)/\sqrt{2}$ made
of large amplitude coherent states, records of different observers 
performing measurements on the environment in a basis correlated with the pointer basis 
will eventually fully agree (as shown in Figure 1), and the 
average relative purity will be equal to one.
For the case when the most predictable states exist, but are not very predictable and are not
imprinted on the environment (and, in particular, do not commute with the interaction 
Hamiltonian), such as the model of two-level atom
resonance fluorescence, the agreement between observers' guess of the state of the system
may only be partial, and it is even possible
to obtain anti-correlation between measurement records, as in the case of
photo-detection.


\section{Acknowledgments} 

We are grateful to Robin Blume-Kohout, Howard Carmichael, Kurt Jacobs,
Harold Ollivier, Juan Pablo Paz and Howard Wiseman for discussions. 
This research was supported in part by NSA. Moreover
J.D. was supported in part by a KBN grant 2 P03B 092 23.


\appendix


\section{ Derivation of the two-level atom photo- and
          homodyne detection master equations for multiple measurement channels}

Let us assume a two-level atom that interacts with the electromagnetic
field, which we shall consider as the environment. We will split this
environment into different parts $i$, each of which has associated
a detector $i$. For example, $i$ may denote different photon
wave vectors. In the rotating wave approximation, the dipole interaction 
between the atom and the electromagnetic field is

\begin{equation}
V(t)=i \sum_{i} ( b^{\dagger}_{i} c - 
c^{\dagger} b_{i} ) ,
\end{equation}
where $b_{i}$ and $c$ are annihilation operators for photons and the atom,
respectively. At every instant of time $t$, a new part of the environment
is interacting with the system. Indeed, a localized
photon wave packet arrives at the atom, interacts with it, and then
flies away. Subsequently, a new wave packet performs the same process.
Imagine that at a given instant of time $t$ the combined state of the atom and
the field is $R(t) = \rho(t) \otimes \mu$, where $\rho$ is the density matrix
for the atom and $\mu$ is that for the field, which we asumme to be in vacuum,
$\mu= \otimes_{i} ~_{i}|0 \rangle \langle 0 |_{i}$. This series of idealizations
are called the quantum Markov approximation.

The evolution operator
for a time interval $dt$ is $U(t,t+dt)=\exp[\sum_{i} 
(dB^{\dagger}_{i} c - c^{\dagger} dB_{i})]$, where 
$dB_{i}(t) = b_{i}(t) dt$ has commutation relations

\begin{equation}
[ dB_{i}(t),dB^{\dagger}_{j}(t)] = \delta_{ij} dt
\end{equation}
that follow from the (singular) commutation relations 
$[db_{i}(t),db^{\dagger}_{j}(t')] = 
\delta_{ij} \delta(t-t') $.
The above commutation relation is of order $dt$ instead of $dt^2$ \cite{wisemanPhD}, as one might
have naively expected. For this reason an expansion 
to first order in $dt$ of 
the evolution operator requires a second order expansion in terms of
$dB_{i}$ and $dB^{\dagger}_{i}$. When ones discards all the 
information contained in the environment (which is then traced out) 
one gets an unconditional master equation for the system

\begin{equation}
d\rho = - i dt [\omega \sigma_x, \rho] + dt \sum_{i}
(c \rho c^{\dagger} - \frac{1}{2} c^{\dagger} c \rho -
\frac{1}{2} \rho c^{\dagger} c ) .
\label{umeapp}
\end{equation}
The sum over $i$ just re-scales the spontaneous emission rate of the atom.
In the following we shall absorb that rescaling in a redefinition of time and
set the spontaneous emission rate to one. 

If the measurements on the environment are not ignored but kept, the evolution
of the system is conditioned upon them. In the case of photo-detection, for most
of the time intervals no photons are detected. In this case of null results
the density matrix of the system evolves according to

\begin{eqnarray}
d \rho_{{\rm zero}}(t) &=& O_{{\rm zero}} \rho(t) - 
\rho(t) {\rm Tr} ( O_{{\rm zero}} \rho(t) ) , \\
O_{{\rm zero}} \rho(t) &=& dt \left( -i [H,\rho] - \frac{1}{2} \sum_{i}
\{ c^{\dagger} c, \rho \} 
\right) ,
\label{ap3ume}
\end{eqnarray}
which is so constructed as to conserve the trace of $\rho$ under the time
evolution. When a photon is measured by any of the detectors, the system
discontinuously jumps to the ground state of the atom

\begin{eqnarray}
d \rho_{{\rm one}}(t) &=& \sum_{i} dN_{i} \left(
\frac{O_{{\rm one}} \rho}{\overline{dN_{i}}} - f_{i} \right) ,\\
O_{{\rm one}} \rho(t) &=& \eta_{i} c \rho c^{\dagger} dt .
\end{eqnarray}
Here the increments $dN_{i} \in \{0,1\}$ are dichotomic stochastic
processes with averages $\overline{dN_{i}(\rho)}=\eta_{i} dt
{\rm Tr} [\rho c^{\dagger} c]$, $\eta_{i}$ denotes the fraction of the
environment measured by detector $i$, and $f_{i}$ is such that
two conditions must be satisfied: 1) when $d \rho =d \rho_{{\rm zero}} + d \rho_{{\rm one}}$
is averaged over all records $i$, it must reduce to the unconditional
master equation, and ;2) ${\rm Tr} [d\rho_{{\rm zero}} + d \rho_{{\rm one}}] =0$. 
It then follows
that $f_{i}=-\eta_{i}\rho {\rm Tr}(O_{{\rm zero}} \rho) / 
\overline{dN_{i}}$.
Finally we get the super-observer master equation for photo-detection

\begin{equation}
d\rho=- i dt \left[ \omega \sigma_x , \rho \right] +
 dt \left( c \rho c^{\dagger} -
 \frac{1}{2} c^{\dagger}c \rho - \frac{1}{2} \rho c^{\dagger}c  \right) + 
 \sum_{i} \left( dN_{i}-\overline{dN_{i}(\rho)} \right) \
\left( \frac{c \rho c^{\dagger} }
            { {\rm Tr} [ c \rho c^{\dagger} ]  } -\rho
\right) .
\end{equation}

The super-observer unconditional master equation (\ref{umeapp}) 
is invariant under the 
transformation $c \rightarrow c+\gamma_{i}$ and 
$H \rightarrow H -(i/2) \sum_{i} (\gamma^{\star}_{i} c -
\gamma_{i} c^{\dagger})$, 
where $\gamma_{i}$ is a complex number \cite{wisemanPhD}. This symmetry
is helpful for deriving other unravelings of the unconditional master equation,
for example the one corresponding to homodyne detection. In this case 
$\gamma_{i}$ represents the coherent amplitude of the classical field
of the local oscillator $i$. Introducing this symmetry into the
photo-detector master equation one immediately obtains the homodyne 
master equation Eq. (\ref{HOMCMEO}).


\section{Calculation of the distribution of waiting times}

In this appendix we calculate the distribution of waiting times  $f_{{\rm wait}}(\tau)$
for the model of resonance fluorescence from a two-level atom subjected 
to direct photo-detection. It is given
by Eq. (\ref{tildap}).

\begin{equation}
f_{{\rm wait}}(\tau) \equiv \sum_{n=0}^{\infty}
  \int_0^{\tau} dt_1 
  \int_{t_1}^{\tau} dt_2 \dots 
  \int_{t_{n-1}}^{\tau} dt_n\;
  D_n(t_1,\dots,t_n,\tau)
  \frac{\eta_1}{2}[ 1 + Z(\tau-t_n) ],
\end{equation}
where we recall that 
\begin{equation}
Z(\tau-t_n) = - e^{-\frac{3}{4} (1-\eta) (\tau-t_n)} 
\cos 2 \omega (\tau-t_n) , 
\end{equation}
and that $D_n$ is  

\begin{equation}
D_n(t_1,\ldots,t_n,\tau) = 
e^{-\frac{\eta}{2} (\tau-t_n)} \prod_{j=1}^n
\frac{\eta_2}{2} e^{-\frac{\eta}{2} (t_j-t_{j-1})}
\left[
1- e^{-\frac{3}{4} (1-\eta) (t_j-t_{j-1})} \cos 2 \omega (t_j-t_{j-1})
\right] ,
\end{equation}
where $t_0=0$ is the time of the last detection by observer 1. Inserting
this
equation into the previous one, we see that when doing the $n$ time
integrals
only two terms will survive: one that stems from the product of all the 1's
in $D_n$, and another coming from the products of all the cosines (which
will therefore contain factors of the form $\cos^2(2 \omega t_j)$). All
other terms in the expansion of the product in $D_n$ will vanish upon 
integration. In the $\omega \gg 1$ limit we can replace $\cos^2(2 \omega t_j)$
by $1/2$. We then get

\begin{eqnarray}
f_{{\rm wait}}(\tau) & \approx & \frac{\eta_1}{2} e^{-\frac{\eta \tau}{2}}
\sum_{n=0}^{\infty}
\int_0^{\tau} dt_1 \ldots \int_{t_{n-1}}^{\tau} dt_n  
\left[
\left( \frac{\eta_2}{2} \right)^n - 
\left( -\frac{\eta_2}{4}\right)^n 
e^{-\frac{3}{4} (1-\eta) \tau} \cos(2 \omega \tau)
\right] \nonumber \\
&= & \frac{\eta_1}{2} e^{-\frac{\eta \tau}{2}} \sum_{n=0}^{\infty}
\left[ \frac{1}{n!} \left(\frac{\eta_2 t}{2}\right)^n -
       \frac{1}{n!} \left(-\frac{\eta_2 t}{4}\right)^n
   e^{-\frac{3}{4} (1-\eta) \tau} \cos(2 \omega \tau) \right] \nonumber \\
&=& \frac{\eta_1}{2} e^{-\frac{\eta_1}{2} \tau} + 
{\cal O} \left( \frac{1}{\omega} \right) .
\end{eqnarray}


\section{Calculation of the relative purity for the two-level atom with
homodyne detection}

In this appendix we derive Eq. (\ref{over2level}) for the stationary
value of the average relative purity between two measurement channels
for the model of resonance fluorescence from a two-level atom subjected 
to homodyne detection.

We must calculate the different terms of Eq. (\ref{o2level}). 
Using Eq. (\ref{sol2level}) we have

\begin{equation}
\overline{ x^{(1)}_{i}(t) x^{(2)}_{j}(t) } =
\eta_{j} \int_{-\infty}^t d\tau e^{-\frac{1}{2}(t-\tau)}
\left(
\cos^2\phi_{j} 
\overline{ x^{(1)}_{i}(t) x^{(1)}_{i}(\tau)} -
\sin\phi_{j} \cos\phi_{j}
\overline{ x^{(1)}_{i}(t) y^{(1)}_{i}(\tau)}
\right).
\end{equation}
Using Eq. (\ref{solution}) and that 
$\overline{ dW_{i} dW_{j}} = \delta_{ij} dt$, it is easy
to show that 
$\overline{ x^{(1)}_{i}(t) y^{(1)}_{i}(\tau)}={\cal O}(1/\omega)$, 
so we can
discard that term in the previous equation. Also,
$\overline{ x^{(1)}_{i}(t) x^{(1)}_{i}(\tau)}=
\eta_{i} \cos^2\phi_{i} \exp(-(t-\tau)/2)$. Hence

\begin{equation}
\overline{ x^{(1)}_{i}(t) x^{(2)}_{j}(t) } = 
\eta_{i} \eta_{j} \cos^2\phi_{i} \cos^2\phi_{j} .
\end{equation}
Also, $\overline{ x^{(1)}_{j}(t) x^{(2)}_{i}(t) }$,
which obtains from the interchange $i \leftrightarrow j$, is the
same. On the other hand, 
\begin{equation}
\overline{ y_{i}^{(1)} y_{j}^{(2)} + z_{i}^{(1)}
z_{j}^{(2)}} = \frac{1}{2}
\overline{ (y_{i}^{(1)} + i z_{i}^{(1)})
(y_{j}^{(2)} - i z_{j}^{(2)}) } + {\rm h.c.} 
\end{equation}
To calculate this noise average, we make use of 
Eqs.(\ref{solu},\ref{sol2level}), and 

\begin{eqnarray}
\overline{ \theta_{i}(\tau) x_{i}^{(1)}(\tau') } &=&
\sqrt{\eta_{i}} \cos\phi_{i} e^{-\frac{1}{2} (\tau-\tau')}
\theta(\tau-\tau') , \nonumber \\
\overline{ \theta_{i}(\tau) y_{i}^{(1)}(\tau')} &=&
- 2 \sqrt{\eta_{i}} \sin\phi_{i} e^{-\frac{3}{4}(\tau'-\tau)}
\theta(\tau-\tau') \cos 2\omega (\tau-\tau') .
\end{eqnarray}
where the stochastic noises $\theta_{i}$ are defined 
as $dW_{i}= \theta_{i} dt$, and
$\theta(\tau)$ is the step function. Performing the necessary time 
integrations and discarding 
${\cal O}(1/\omega)$ terms, we get

\begin{equation}
\overline{ y_{i}^{(1)} y_{j}^{(2)} + z_{i}^{(1)}
z_{j}^{(2)}} = \frac{4}{9} \eta_{i} \eta_{j}
\sin^2 \phi_{i} \sin^2 \phi_{j} .
\end{equation}
Finally, the average relative purity between the two single observer density matrices 
reads

\begin{equation}
O_{ij} = \frac{1}{2} + \eta_{i} \eta_{j}
\left[
\cos^2\phi_{i} \cos^2\phi_{j} + \frac{4}{9}
\sin^2\phi_{i} \sin^2\phi_{j}
\right] .
\end{equation}


\end{document}